\documentclass[twocolumn,aps,prc,showpacs,superscriptaddress,preprintnumbers,floatfix,nofootinbib]{revtex4}
\usepackage{epsfig,graphics}
\usepackage{graphicx}
\usepackage{dcolumn}
\usepackage{bm}
\usepackage{amsmath}
\usepackage[usenames]{color}
\usepackage{ulem} 
\usepackage[colorlinks,linkcolor=blue,urlcolor=blue,citecolor=blue]{hyperref}

\begin{document}

\title{Initial partonic eccentricity fluctuations in a multi-phase transport model}

\author{ L. Ma}
\affiliation{Shanghai Institute of Applied Physics, Chinese
Academy of Sciences, Shanghai 201800, China}
 \affiliation{University of  Chinese Academy of Sciences, Beijing 100049, China}
\author{G. L. Ma}
\email[]{glma@sinap.ac.cn}
\affiliation{Shanghai Institute of Applied Physics, Chinese
Academy of Sciences, Shanghai 201800, China}
\author{Y. G. Ma}
\email[]{ygma@sinap.ac.cn}
\affiliation{Shanghai Institute of Applied Physics, Chinese
Academy of Sciences, Shanghai 201800, China}
\affiliation{ShanghaiTech University, Shanghai 200031, China}

\date{}

\begin{abstract}

Initial partonic eccentricities in Au+Au collisions at center-of-mass energy $\sqrt{s_{NN}}$ = 200 GeV are investigated using a multi-phase transport model with string melting scenario. The initial eccentricities in different order of harmonics are studied using participant and cumulant definitions. Eccentricity in terms of second-, fourth- and sixth order cumulants as a function of number of participant nucleons are compared systematically with the traditional participant definition. The ratio of the cumulant eccentricities $\varepsilon\left\{4\right\}/\varepsilon\left\{2\right\}$ and $\varepsilon\left\{6\right\}/\varepsilon\left\{4\right\}$ are studied in comparison with the ratio of the corresponding flow harmonics. The conversion coefficients ($v_n/\varepsilon_n$) are explored up to fourth order harmonic based on cumulant method. Furthermore, studies on transverse momentum ($p_T$) and pseudo-rapidity ($\eta$) dependencies of eccentricities and their fluctuations are presented. As in ideal hydrodynamics initial eccentricities are expected to be closely related to the final flow harmonics in relativistic heavy-ion collisions, studies of the fluctuating initial condition in the AMPT model will shed light on the tomography properties of the initial source geometry.

\end{abstract}


\maketitle

\section{Introduction}

In ultra-high energy heavy-ion collisions, the pressure gradients in the overlap zone is large enough to translate the initial coordinate space anisotropy to the final state momentum space anisotropy which can be experimentally observed as anisotropic flow.  Anisotropic flow as a typical collective behavior of emitted particles  has been proved to be a good observable to study the new matter in relativistic heavy-ion collisions providing information on equation-of-state and the transport properties of the matter created~\cite{Ollitrault1992,Voloshin2008}. One of the most striking experimental results ever obtained in relativistic heavy-ion collisions is the strong elliptic flow ($v_{2}$). The fluid-like behavior of matter created in the early stage leads to the conclusion that the quark gluon plasma is like a nearly perfect liquid~\cite{Kolb1999,Ackermann2001,Teaney2001,Romatschke2007,Song2011,Ko}.  Due to the fluid-like properties of elliptic flow $v_{2}$, hydrodynamic models have been widely used to make predictions and it was suggested that the final state anisotropy inherits information from initial state and carries additional information of the system evolution~\cite{Ulrich2005,Gale2013,Qiu20111441}. Thus, measurement of elliptic flow coefficient $v_{2}$ provides essential information about the hot and dense matter created in relativistic heavy-ion collisions.

Besides the elliptic flow $v_{2}$ measurement, higher-order harmonic flow coefficients defined as $v_{n}$ (n = 3,4,5) drawn much more attention in both experiment and model studies in recent years as higher harmonics are suggested to be sensitive to the initial partonic dynamics~\cite{STAR2004flow,STAR2013flow,PHENIX2011flow,Chen2004,Han2011,Phenix2}. The importance of fluctuations was firstly realized in the simulations with a multi-phase transport (AMPT) model, showing that the fluctuating initial source geometry transfers to the final momentum space during the system expansion leading to non-zero higher odd order harmonic flow coefficients~\cite{Alver201081}. It was also realized that higher harmonics like triangular flow $v_3$ is particularly sensitive not only to the initial condition but also to the shear viscosity $\eta$/s which reflects the properties of the source in the early stage~\cite{Schenke2011,Schenke2012}. Studies also suggest that the elliptic flow and higher harmonic flow fluctuations on the event-by-event basis elucidate both the system dynamics and new phenomena occuring in the very beginning of collisions~\cite{Andrade2006,Petersen2010,Ma2014}. Experimental measurements on an event-by-event basis were done for the elliptic flow $v_2$ fluctuation study in Au+Au collisions at $\sqrt{s_{NN}}$ = 200 GeV in PHOBOS and STAR experiments~\cite{Sorensen2007,Alver200734,Alver2010104,Agakishiev2011}. It suggests that the close correlation between anisotropic flow fluctuation and the fluctuations of the initial source geometry  carry important information of the viscosity and other properties of the matter created in heavy-ion collisions~\cite{Holopainen2010,Alver201082}. 

Significant attention has been paid to the studies of initial geometry fluctuation effect on the final flow observables~\cite{Miller2003,DerradideSouza2011,Ma2011106,Wang2014,WangJ2}. The essential role of the collision geometry was realized when one looks into the flow harmonics of different collision systems scaled by initial eccentricities~\cite{Alver200798}.  Phenomena observed strongly suggest that the partonic participant eccentricity is responsible for the development of the final anisotropic flow.  Higher order eccentricities are also suggested closely related to the final higher order harmionic flow. The triangular flow ($v_3$) and higher harmonics are suggested arise from event-by-event initial fluctuations which lead to finite value even in most central collisions. Thus, the study of initial eccentricity and fluctuation is crucial for understanding of final flow and flow fluctuation~\cite{Drescher2007,Broniowski2007}. Therefore, the study of the event-by-event flow response to the initial eccentricity in model simulation is important for a quantitative study of the source evolution properties in relativistic heavy-ion collisions.

In this paper, we present specific discussion on source eccentricity and their fluctuation properties in the initial partonic stage of the high energy heavy-ion collision using a multi-phase transport (AMPT) model. Systematic comparisons are made between cumulant eccentrities and participant eccentrities. Centrality, pseudo-rapidity and transverse momentum dependences of higher order harmonics are studied in model simulations providing tomographic pictures of the source profile. The results are expected to give additional constraints on the initial source condition. This paper is organized as follows: In Sec. II, a multi-phase transport (AMPT) model is introduced briefly. In Sec. III. results and discussions are presented. Last section is a brief summary.

\section{Brief description of AMPT model}

A multi-phase transport model (AMPT)~\cite{Zhang2000} is an useful model for investigating reaction dynamics in relativistic heavy-ion collision. There are two versions with different scenarios - the default version and the string melting version both of which consist of four main components: the initial condition, partonic interactions, hadronization, and hadronic interactions.

In the initial stage, the phase space distributions of minijet partons and soft string excitations are included which come from the Heavy Ion Jet Interaction Generator (HIJING) model~\cite{Wang1991}. Multiple scatterings lead to fluctuations in local parton number density and hot spots from both soft and hard interactions which are proportional to local transverse density of participant nucleons. In AMPT string melting version, both excited strings and minijet partons are decomposed into partons. Scatterings among partons are then treated according to a parton cascade model - Zhang's parton cascade (ZPC) model which includes parton-parton elastic scattering with cross sections obtained from the theory calculations~\cite{Zhang1997}.  After partons stop interacting with each other, a simple quark coalescence model is used to combine partons into hadrons. Partonic matter is then turned into  hadronic matter and the subsequential hadronic interactions are modelled using a relativistic transport model (ART) including both elastic and inelastic scattering descriptions for baryon-baryon, baryon-meson and meson-meson interactions~\cite{Lin2005}. 

In the ZPC parton cascade model, the differential scattering cross section for partons is defined as:

\begin{equation}
\frac{d\sigma_{p}}{dt}=\frac{9\pi\alpha^{2}_{s}}{2}(1+\frac{\mu^{2}}{s})\frac{1}{(t-\mu^{2})^{2}},
\label{q1}
\end{equation}
where $\alpha_{s}$ = 0.47 is the strong coupling constant, $s$ and $t$ are the usual Mandelstam variables and $\mu$ is the screening mass in partonic matter. Studies show that a multiphase transport model with a string melting scenario gives better description of experimental measurements of anisotropic flow harmonics. With properly choosing parton scattering cross section, data on harmonic flow of charged hadrons measured from experiments for Au+Au collisions at 200GeV can be approximately reproduced~\cite{Han2011}. Recent studies also showed that by changing input parameters, AMPT could quantitatively describe the centrality dependence of elliptic flow and triangular flow in Au + Au ~\cite{Xu2011} as well as the vector mesons in p + p and  d + Au systems \cite{Ye,Xu}.

In this work, we use AMPT string melting version to simulate Au+Au collisions. Our default sample of simulated events for Au+Au collisions at the center-of-mass energy of 200 GeV are generated with a parton cross section of 3mb. But we will compare 3 mb with 10 mb when we study the effect of parton cross section. We make a description of Au+Au collision at 200GeV with AMPT by using parameter set: a=2.2, b=0.5  (GeV$^{-2}$) in the Lund string fragmentation function as shown in Ref.~\cite{Lin2001}. Particularly, the hadronic scattering effect and resonance decay effect on the harmonic flow evolution are both taken into account in the model simulation.

Table.~\ref{table1} shows different centrality classes divided for the simulation samples. The mean number of participant nucleons and corresponding impact parameter for each centrality bin are also shown in the table. In this paper, centrality dependence of all kinds of observables can be measured as a function of mean number of participant nucleons.

\begin{table}[htbp]
\caption{ Centrality classes of AMPT events in Au+Au collisions at $\sqrt{s_{NN}}$ = 200 GeV. }
\label{table1}
\centering
\begin{tabular}{p{60pt}p{100pt}p{60pt}}
\hline
\hline
Centrality & Impact Parameter Range (fm) & $\left\langle{Npart}\right\rangle$\\
\hline
 0$\%$ - 10$\%$ & 0.00 - 4.42      & 345.8$\pm$0.1 \\
10$\%$ - 20$\%$ & 4.42 - 6.25      & 263.5$\pm$0.1 \\
20$\%$ - 30$\%$ & 6.25 - 7.65      & 198.2$\pm$0.0 \\
30$\%$ - 40$\%$ & 7.65 - 8.83      & 146.8$\pm$0.1 \\
40$\%$ - 50$\%$ & 8.83 - 9.88      & 106.1$\pm$0.0 \\
50$\%$ - 60$\%$ & 9.88 - 10.82     & 73.8$\pm$0.1 \\
60$\%$ - 70$\%$ & 10.82 - 11.68    & 48.8$\pm$0.2 \\
\hline
\hline
\end{tabular}
\end{table}

\section{Results and Discussion}

\subsection{ Initial eccentricity and eccentricity fluctuation in partonic stage of AMPT model}

Experimental measurements of flow coefficients $v_{n}$ could be affected by event-by-event fluctuations in the initial geometry. Considering the event-by-event fluctuation effect, harmonic flow $v_{n}$ was proposed to calculate with respect to the participant plane angle $\psi_{n}\left\{part\right\}$ under participant coordinate system instead of the traditional reaction plane angle $\psi_{RP}$ in the model simulation~\cite{Voloshin2007695}. The above method for the calculation of $v_n$ is referred to as participant plane method which has been widely used for flow calculations in different models~\cite{DerradideSouza2011}. The participant plane is defined as

\begin{equation}
\psi_{n}\left\{part\right\}  = \frac{1}{n}\left[ \arctan\frac{\left\langle {r^n \sin
(n\varphi)} \right\rangle}{\left\langle {r^n \cos (n\varphi)}
\right\rangle} + \pi \right],
 \label{q2}
\end{equation}
where $n$ denotes the $n$th-order participant plane,  $r$ and $\varphi$ are the position and azimuthal angle of each parton in AMPT initial stage and the average $\langle \cdots\rangle$ denotes density weighted average.  Harmonic flow coefficients with respect to the participant plane are defined as

\begin{equation}
v_{n}\left\{part\right\} = \left\langle cos[n(\phi-\psi_{n}\left\{part\right\})] \right\rangle,
\label{q3}
\end{equation}
where $\phi$ is azimuthal angle of final particle, and the average $\langle \cdots\rangle$ denotes particle average.

Similar to the harmonic flow coefficient, different definitions of the initial anisotropy coefficients are described in Ref.~\cite{Qiu201184}. The one referred to as "participant eccentricity" which characterizes the initial state through the event-by-event distribution of the participant nucleons or partons has been found to be crucial for understanding the initial properties~\cite{Alver200798}. The participant eccentricity for initial elliptic anisotropy is given by

\begin{equation}
\varepsilon_{2}\left\{part\right\} = \frac{\sqrt{(\sigma^{2}_{y}-\sigma^{2}_{x})^{2}+4(\sigma_{xy})^{2}}}{\sigma^{2}_{y}+\sigma^{2}_{x}},
\label{q4}
\end{equation}
where $\sigma_{x}$,$\sigma_{y}$,$\sigma_{xy}$ ,are the event-by-event variances of the participant nucleon or parton distributions along the transverse directions $x$ and $y$. When transforming the coordinate system to the center-of-mass frame of the participating nucleons, a genelrized definition of $\varepsilon_{n}\left\{part\right\}$  n-th order participant eccentricities is in the form~\cite{Petersen2012}

\begin{equation}
\varepsilon _{n}\left\{part\right\}  = \frac{{\sqrt {\left\langle {r^{n} \cos (n\varphi)} \right\rangle ^2 + \left\langle {r^{n} \sin (n\varphi)}
\right\rangle ^2 } }}{{\left\langle {r^{n} } \right\rangle }},
\label{q5}
\end{equation}
where $r$ and $\varphi$ are the same definitions as for participant plane.  Such definition does not make reference to the direction of the impact parameter vector and instead characterizes the eccentricity through the distribution of participant nucleons or partons which naturally contains event-by-event fluctuation effect. We simply take this as participant definition or participant method. 

As indicated by Ref.~\cite{Voloshin2006,Bhalerao2006}, under the assumption that $v_{2}$ from participant plane method $v _{2}\left\{part\right\}$ is proportional to $\varepsilon _{2}\left\{part\right\}$, scaling properties are expected to hold for even higher harmonics. Similar to flow harmonics, it is proposed that initial eccentricity can be quantified by cumulants of $\varepsilon _{n}\left\{part\right\}$ ~\cite{Qiu201184}. The definitions of second, fourth, sixth order cumulant of $\varepsilon _{n}\left\{part\right\}$ are in the form

\begin{equation}
\begin{split}
c_{\varepsilon _{n}\left\{part\right\}}\left\{2\right\}  &= \left\langle{\varepsilon^{2}_{n}\left\{part\right\}}\right\rangle,\\
c_{\varepsilon _{n}\left\{part\right\}}\left\{4\right\}  &= \left\langle{\varepsilon^{4}_{n}\left\{part\right\}}\right\rangle-2\left\langle{\varepsilon^{2}_{n}\left\{part\right\}}\right\rangle^{2},\\
c_{\varepsilon _{n}\left\{part\right\}}\left\{6\right\}  &= \left\langle{\varepsilon^{6}_{n}\left\{part\right\}}\right\rangle-9\left\langle{\varepsilon^{2}_{n}\left\{part\right\}}\right\rangle\left\langle{\varepsilon^{4}_{n}\left\{part\right\}}\right\rangle\\
&+12\left\langle{\varepsilon^{2}_{n}\left\{part\right\}}\right\rangle^{3}.
\end{split}
\label{q6}
\end{equation}

For the definitions~(\ref{q6}), the cumulant definitions here follow the regular way of cumulant flow definitions for two, four and six particle azimuthal correlations as in Ref.~\cite{Miller2003}. The corresponding eccentricities defined by cumulants are written as

\begin{equation}
\begin{split}
&\varepsilon^{RC} _{n}\left\{2\right\} = \sqrt{ c_{\varepsilon _{n}\left\{part\right\}}\left\{2\right\} },\\
&\varepsilon^{RC} _{n}\left\{4\right\} = ( -c_{\varepsilon _{n}\left\{part\right\}}\left\{4\right\} )^{1/4},\\
&\varepsilon^{RC} _{n}\left\{6\right\} = ( c_{\varepsilon _{n}\left\{part\right\}}\left\{6\right\}/4 )^{1/6}.
\end{split}
\label{q7}
\end{equation}

Here, we use superscript "RC" to denote the definition of the regular cumulant commonly used in many studies~\cite{DerradideSouza2011,Agakishiev2011}. Experimentally, as the initial state in heavy-ion collisions is not accessible, the participant plane method is not applicable. Instead, particle correlation method was proposed for flow study via measurement of correlation of final particles without assumimg a certain participant plane. In recent years, a multi-particle cumulants method called Q-cumulant or direct cumulant method was proposed and widely used in both model and experimental studies~\cite{Bilandzic2011,DerradideSouza2011,Zhou2015,Abelev2014,Ma2014}. This method uses the Q-vector to calculate directly the multiparticle cumulants. The Q-vector is defined as

\begin{equation}
Q_{n}=\sum_{i=1}^{M}e^{in\phi_{i}},
\label{q8}
\end{equation}
where $\phi_i$ is the azimuthal angle in the momentum space of the final particles. The derivation of the expressions for higher order cumulants is straightforward and the two-, four- and six- particle cumulants can be written as

\begin{equation}
\begin{split}
\langle 2 \rangle &=\langle e^{in(\phi_{1}-\phi_{2})}\rangle =\frac{|Q_{n}|^{2}-M}{M(M-1)},\\
\langle 4 \rangle &=\langle e^{in(\phi_{1}+\phi_{2}-\phi_{3}-\phi_{4})}\rangle \\
&=[|Q_{n}|^{4}+|Q_{2n}|^{2}-2Re[Q_{2n}Q^{*}_{n}Q^{*}_{n}]-2[2(M-2)|Q_{n}|^2\\
&-M(M-3)]]/[M(M-1)(M-2)(M-3)],\\
\langle 6 \rangle &=\langle e^{in(\phi_{1}+\phi_{2}+\phi_{3}-\phi_{4}-\phi_{5}-\phi_{6})}\rangle \\
&=[|Q_{n}|^{6}+9|Q_{2n}|^{2}|Q_{n}|^{2}-6Re(Q_{2n}Q_{n}Q^{*}_{n}Q^{*}_{n}Q^{*}_{n})\\
&+4Re(Q_{3n}Q^{*}_{n}Q^{*}_{n}Q^{*}_{n})-12Re(Q_{3n}Q^{*}_{2n}Q^{*}_{n})+18(M-4)\\
&Re(Q_{2n}Q^{*}_{n}Q^{*}_{n})+4|Q_{3n}|^{2}-9(M-4)(|Q_{n}|^{4}+|Q_{2n}|^{2})\\
&+18(M-2)(M-5)|Q_{n}|^{2}-6M(M-4)(M-5)]\\
&/[M(M-1)(M-2)(M-3)(M-4)(M-5)]
\end{split}
\label{q9}
\end{equation}

Then, the second- and fourth-order cumulants on event average can be given by:
\begin{equation}
\begin{split}
&c_{n}\left\{2\right\}=\langle \langle 2\rangle \rangle,\\
&c_{n}\left\{4\right\}=\langle \langle 4\rangle \rangle -2\langle \langle 2\rangle \rangle ^{2},\\
&c_{n}\left\{6\right\}=\langle \langle 6\rangle \rangle -9\langle \langle 2\rangle \rangle \langle \langle 4\rangle \rangle +12\langle \langle 2\rangle \rangle ^{3},
\end{split}
\label{q10}
\end{equation}
where the double brackets denote weighted average of multi-particle correlations. The weights are the total number of combinations from two-, four-, or six-particle correlations, respectively. For flow coefficient with 2-particle cumulant, in order to suppress non-flow from short range correlations, we divide the whole event into two sub-events A and B separated by a pseudo-rapidity gap of 0.3. Then, $\langle 2 \rangle$ in Eq.\ref{q9} is modified to be

\begin{equation}
\langle 2 \rangle_{\Delta\eta} =\frac{Q^{A}_{n}\cdot Q^{B}_{n}}{M^{A}\cdot M^{B}},
\label{q11}
\end{equation}
where $Q^{A}$ and $Q^{B}$ are the flow vectors from sub-event A and B, with $M^{A}$ and $M^{B}$ the corresponding multiplicities.

Then, the harmonic flow $v_{n}$ can be estimated via cumulants (n = 2,3,4..):

\begin{equation}
\begin{split}
&v_{n}\left\{2\right\}=\sqrt{c_{n}\left\{2\right\}},\\
&v_{n}\left\{4\right\}=\sqrt[4]{-c_{n}\left\{4\right\}},\\
&v_{n}\left\{6\right\}=\sqrt[6]{c_{n}\left\{6\right\}/4}
\end{split}
\label{q12}
\end{equation}

Estimations of differential flow (for second- and fourth-order cumulants) can be expressed as:

\begin{equation}
v^{'}_{n}\left \{ 2 \right \}=\frac{d_{n}\left \{ 2 \right \}}{\sqrt{c_{n}\left \{ 2 \right \}}},
v^{'}_{n}\left \{ 4 \right \}=\frac{d_{n}\left \{ 4 \right \}}{-c_{n}\left \{ 4 \right \}^{3/4}}
\label{q13}
\end{equation}
where the $d_{n}\left \{ 2 \right \}$ and $d_{n}\left \{ 4 \right \}$ are the two- and four-particle differential cumulants as defined in Ref.~\cite{Bilandzic2011}.

Cumulant method has been applied very successfully in the studies of harmonic flow coefficients and initial eccentricity in heavy-ion collisions~\cite{STAR200266,Miller2003}. It can be extended to the study of initial-state eccentricity fluctuation which can be in a similar way as flow fluctuation study with cumulant method. The relative fluctuation of $\varepsilon_{n}$ in cumulant definition can be written as

\begin{equation}
R_{\varepsilon_{n}} = \sqrt{\frac{\varepsilon^{2}_{n}\left\{2\right\}-\varepsilon^{2}_{n}\left\{4\right\}}{\varepsilon^{2}_{n}\left\{2\right\}+\varepsilon^{2}_{n}\left\{4\right\}}}.
\label{q14}
\end{equation}

\begin{figure*}[htbp]
\centering
\includegraphics[scale=0.9]{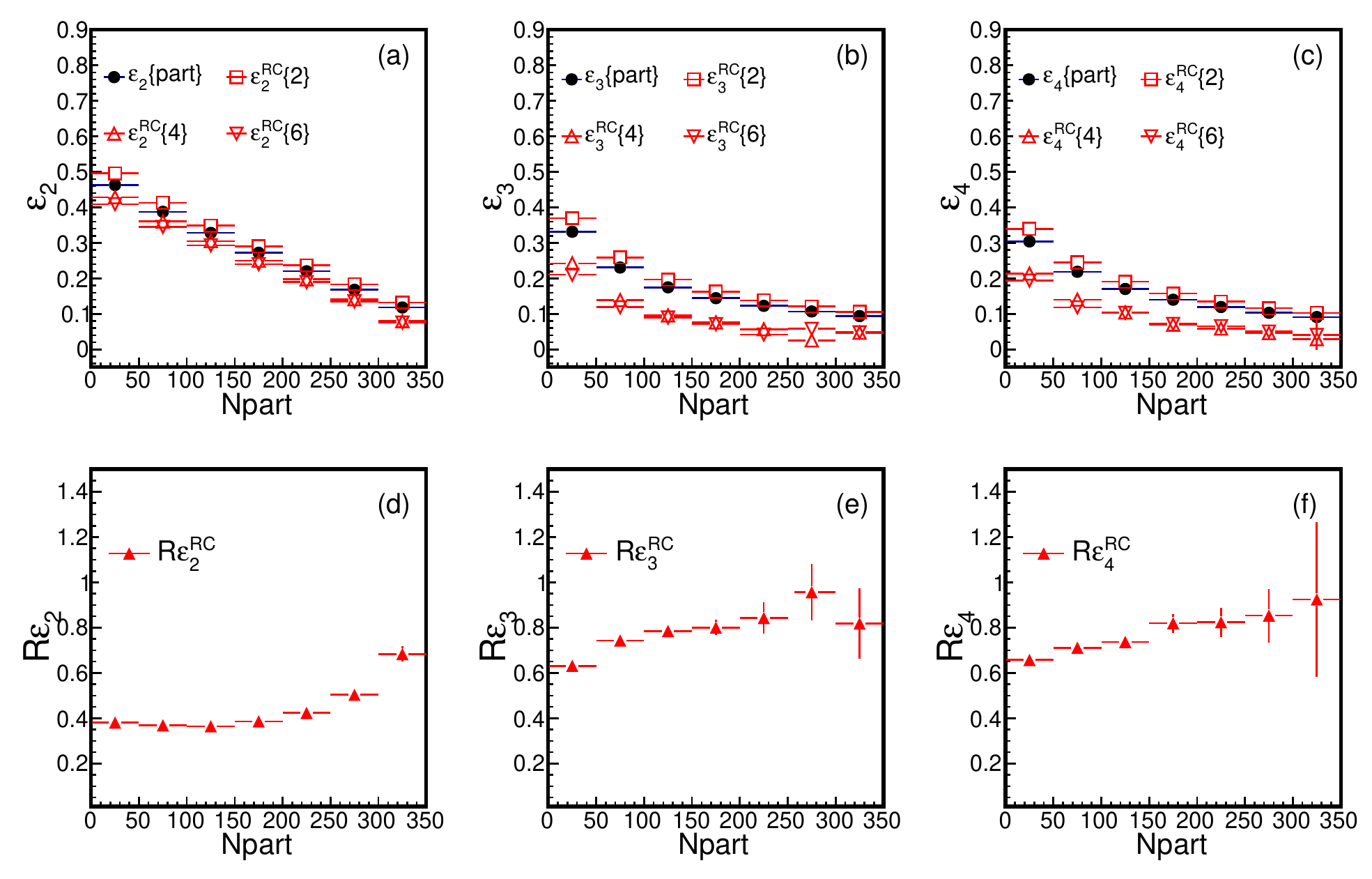}
\caption{(Color online) Initial partonic eccentricity $\varepsilon_{n}$ (n=2,3,4) and their relative fluctuations defined by participant and regular cumulant methods as a function of mean number of participant nucleons $N_{part}$. Eccentricities $\varepsilon _{n}\left\{2\right\}$, $\varepsilon _{n}\left\{4\right\}$ and $\varepsilon _{n}\left\{6\right\}$ defined based on Eq.(\ref{q7}) are denoted as $\varepsilon^{RC}_{n}\left\{k\right\}$(k=2,4,6). Upper panels: Initial partonic eccentricity $\varepsilon_{n}$ in different orders of harmonics. Lower panels: Relative fluctuations of eccentricities in different orders of harmonics defined by Eq.(\ref{q14}).}
\label{f1}
\end{figure*}

It has been argued that the magnitudes and trends of the eccentricities $\varepsilon_{n}$ imply specifically testable predictions for the magnitude and centrality dependence of flow harmonics $v_{n}$~\cite{Lacey2010}. We make a comparison of eccentricities in both cumulant and participant definitions as a function of mean number of participant nucleons $N_{part}$. Upper panels of Figure~\ref{f1}  show $N_{part}$  dependence of two- , four- and six particle cumulant eccentricity $\varepsilon_{n}\left\{2\right\}$, $\varepsilon_{n}\left\{4\right\}$, $\varepsilon_{n}\left\{6\right\}$ and also the participant eccentricity $\varepsilon_{n}\left\{part\right\}$ for different harmonics in Au+Au collision at 200 GeV in the AMPT model. Cumulant eccentricities are defined with regular method ($\varepsilon^{RC}$) from multi-particle correlation of the initial partons in the AMPT initial stage. It is found that $\varepsilon_{n}\left\{part\right\}$ (n=2,3,4) are quantitatively smaller than $\varepsilon_{n}\left\{2\right\}$ and larger than $\varepsilon_{n}\left\{k\right\}$ (k=4,6) over the whole centrality range. $\varepsilon_{n}$ from different definitions show similar trend as a function of mean number of participant nucleons. 

In the lower panel of Figure~\ref{f1}, we plot the relative fluctuations of initial partonic eccentricities in different orders of harmonics as a function of $N_{part}$. In comparison, fluctuation of elliptic eccentricity from regular cumulant definition exhibits clear dependence on the centrality while higher order eccentricity fluctuations show little centrality dependence. Fluctuations of eccentricities $R_{\varepsilon^{RC}_{n}}$(n=2,3,4) are systematically larger for central collisions than non-central collisions. For higher order harmonics, fluctuations $R_{\varepsilon^{RC}_{n}}$ (n=3,4) are larger than $R_{\varepsilon^{RC}_{2}}$ for mid-central or peripheral collisions but comparable in magnitude for central collisions.

\begin{figure*}[htbp]
\centering
\includegraphics[scale=0.9]{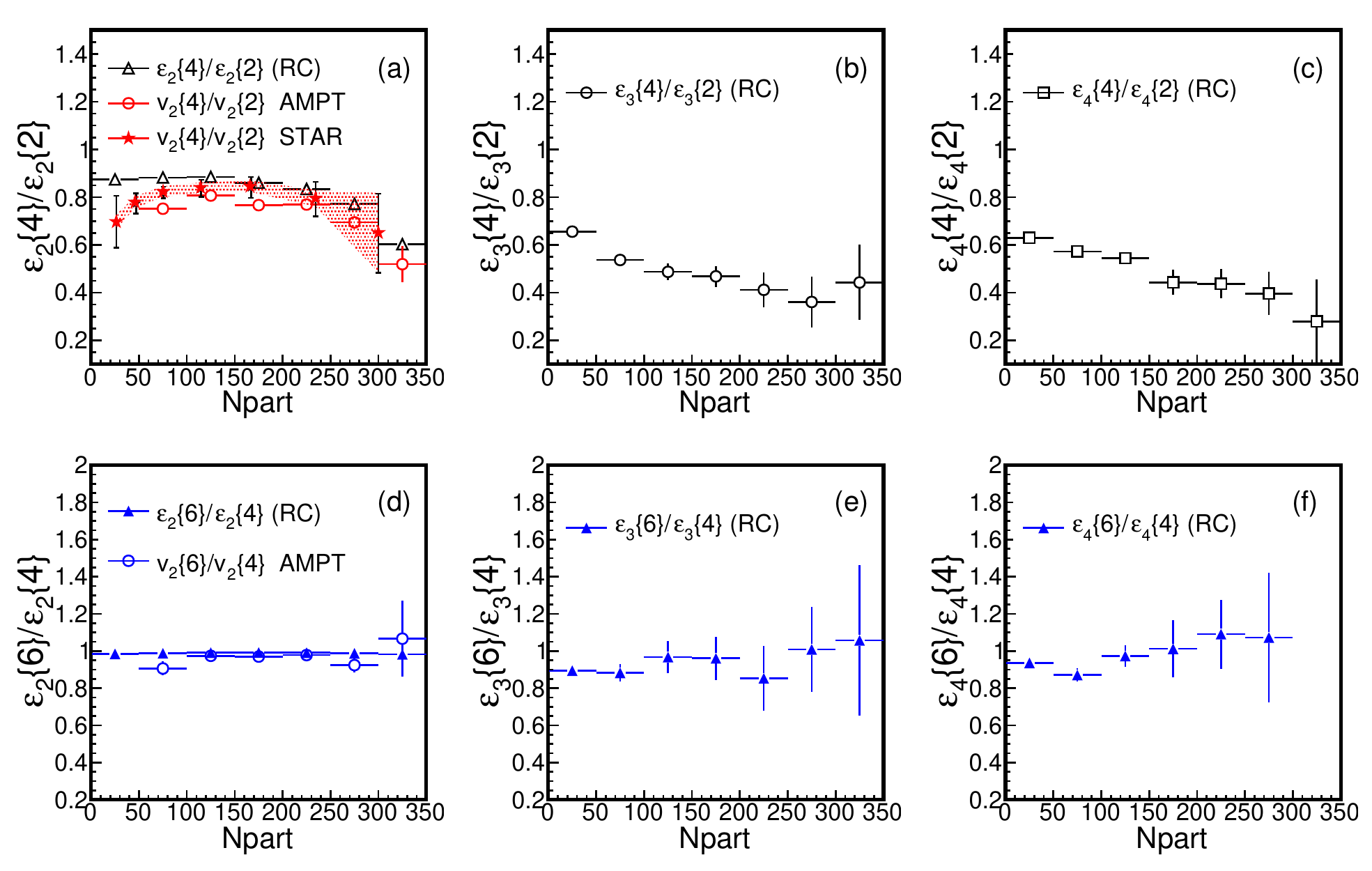}
\caption{(Color online) Cumulant ratios $\varepsilon_{n}\left\{4\right\}/\varepsilon_{n}\left\{2\right\}$ and $\varepsilon_{n}\left\{6\right\}/\varepsilon_{n}\left\{4\right\}$ (n=2,3,4) as a function of $N_{part}$. The eccentricities are defined with regular cumulant method form Eq.\ref{q7} (denote as RC). Upper panels: Cumulants ratio $\varepsilon_{n}\left\{4\right\}/\varepsilon_{n}\left\{2\right\}$ versus $N_{part}$. Results of $v_{n}\left\{4\right\}/v_{n}\left\{2\right\}$ from both experiment measurement with Q-cumulant method and AMPT are also shown for comparison. Lower panels: Cumulants ratio $\varepsilon_{n}\left\{6\right\}/\varepsilon_{n}\left\{4\right\}$ versus $N_{part}$ from cumulant definitions.}
\label{f2}
\end{figure*}

It has been shown that the relative magnitude of $v_{n}\left\{2\right\}$ and $v_{n}\left\{4\right\}$ depends on the fluctuations of $v_{n}$. Assuming that $v_{n}$ is proportional to $\varepsilon_{n}$ on an event-by-event basis, the following equation holds for higher orders (n=2,3,4)

\begin{equation}
\frac{v_{n}\left\{4\right\}}{v_{n}\left\{2\right\}} = \frac{\varepsilon_{n}\left\{4\right\}}{\varepsilon_{n}\left\{2\right\}} = (2-\frac{\langle\varepsilon^{4}_{n}\rangle}{\langle\varepsilon^{2}_{n}\rangle^{2}})^{-1}.
\label{q15}
\end{equation}

Fluctuations of $v_{n}$ are supposed to stem from the fluctuations of $\varepsilon_{n}$~\cite{Bhalerao2011}. Figure~\ref{f2} displays ratios of cumulant eccentricities up to the fourth order in the AMPT model using regular cumulant method. $\varepsilon_{n}\left\{4\right\}/\varepsilon_{n}\left\{2\right\}$ shows a smooth decreasing trend from peripheral collisions to central collisions. The ratio is smaller than unity as expected due to the definition. The smaller eccentricity fluctuation, the closer the ratio to unity. $v_{n}\left\{4\right\}/v_{n}\left\{2\right\}$ from AMPT and experimental flow measurements based on Q-cumulant which scale like the corresponding ratios of eccentricity cumulants are shown in comparison. It is found $\varepsilon_{2}\left\{4\right\}/\varepsilon_{2}\left\{2\right\}$ are roughly equal to the ratio of the flow harmonic $v_{2}\left\{4\right\}/v_{2}\left\{2\right\}$ for non-peripheral collisions. Ratios of six-particle cumulant to four-particle cumulant $\varepsilon_{2}\left\{6\right\}/\varepsilon_{2}\left\{4\right\}$ is roughly equal to unity without seen any centrality dependence which is in consistent with the ratio of the flow harmonic $v_{2}\left\{6\right\}/v_{2}\left\{4\right\}\sim1$. Ratio of higher order harmonics $\varepsilon_{n}\left\{4\right\}/\varepsilon_{n}\left\{2\right\}$ and $\varepsilon_{n}\left\{6\right\}/\varepsilon_{n}\left\{4\right\}$ (n=3,4) are also shown providing additional constraints on the predictions of the ratio of the cumulant flow. Further experimental study of the ratio between cumulant flow harmonics may give access to the initial profile assumming the proportional relation between initial and final anistropies~\cite{Yan2013112,Yan201490}.

Recent theoretical works show increasing interests in longitudinal features of the source created by relativistic heavy-ion collisions~\cite{Pang201286,Pang201552,Bozek2015,Xiao2012}. A model simulation shows that initial-state longitudinal fluctuations for second and third order harmonis survive the collective expansion resulting in a forward-backward asymmetry which propagate to the final stage during the source evolution~\cite{Jia201490}. Experimentally, flow measurements have been extended to study the longitudinal behaviour of flow harmonics~\cite{PHOBOS200572,Adam2016}. Due to the close relation between initial geometry and final flow harmonics, a systematic study of the longitudinal profile of the source is crucial for the understanding of the source evolution.

\begin{figure*}[htbp]
\centering
\includegraphics[scale=0.9]{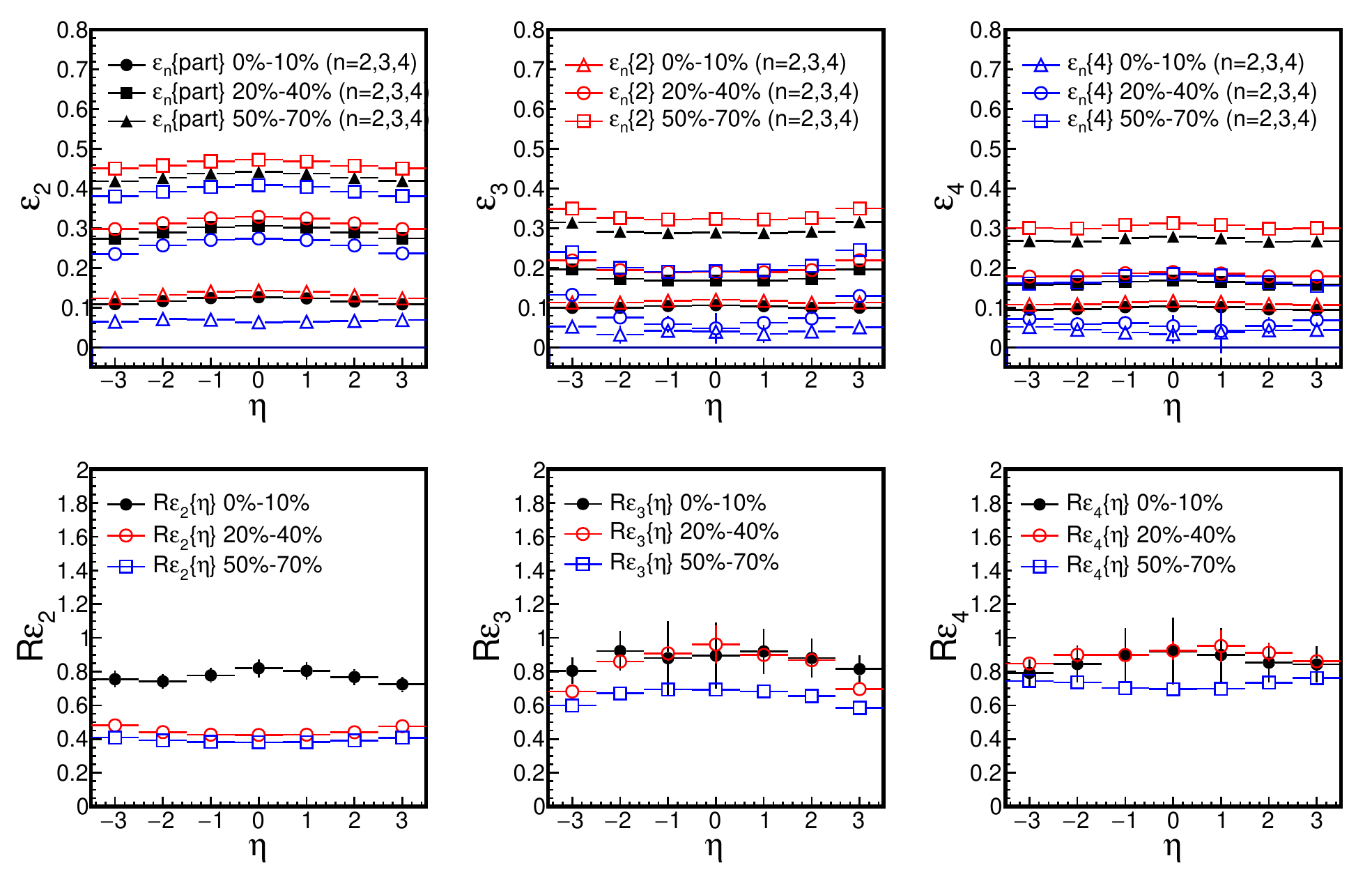}
\caption{(Color online) Eccentricity coefficients $\varepsilon_{n}$ (n=2,3,4) defined with participant method and regular cumulant method as a function of pseudo-rapidity ($\eta$) for the AMPT initial condition. Fluctuation of $\varepsilon_{n}$ is studied up to the fourth order harmonic based on Eq.(\ref{q14}). Upper panel: $\varepsilon_{n}$ versus $\eta$ defined by participant and cumulant method. Lower panel: $\varepsilon_{n}$ fluctuation up to the fourth order as a function of $\eta$. Results are shown for three selected centrality classes in Au+Au collisions at $\sqrt{s_{NN}}$ = 200 GeV.}
\label{f4}
\end{figure*}

We perform here an investigation on the pseudo-rapidity $\eta$ dependence of $\varepsilon_{n}$ in the AMPT partonic stage of Au+Au collisions at 200 GeV. In the upper panels of Figure~\ref{f4}, $\varepsilon_{n}\left\{part\right\}$ are shown as a function of pseudo-rapidity $\eta$ for three different centrality classes, where two- and four- particle cumulant $\varepsilon_{n}$ defined by Eq.(\ref{q7}) are plotted in additional to the participant $\varepsilon_{n}\left\{part\right\}$. Participant and cumulant eccentricity show almost the same trend as a function of $\eta$. Comparing to the results of flow fluctuation in the AMPT calculations as shown in previous study~\cite{Ma2014}, $\varepsilon_{2}(\eta)$ is in a similar trend to $v_{2}(\eta)$ at the same centrality. $\varepsilon_{3}(\eta)$ or $\varepsilon_{4}(\eta)$ shows little $\eta$ dependence which is quite different from corresponding flow harmonic $v_{3}(\eta)$ or $v_{4}(\eta)$. One possible cause might be from the partonic evolution process, but more investigations are needed for the final conclusion. As seen in the lower pannel of Fig.\ref{f4}, relative fluctuation of the participant eccentricity from regular cumulant definition $R_{\varepsilon_{n}}$ (n=2,3,4) show symmetric profile as a function of pseudo-rapidity with tiny $\eta$ dependence for higher order harmonics (n$\geq$3) which is quite similar to the flow fluctuation. As pseudo-rapidity dependences of initial eccentricities reflect the longitudinal features of the created partonic matter, systematic comparison between eccentricity in model simulation and flow harmonic and their fluctuation properties in experiments are necessary to provide a valuable information for comprehensive understanding of the created source.  

Besides investigating pseudo-rapidity($\eta$) dependence of eccentricity, it is also important to check the transverse momentum ($p_{T}$) dependence of initial eccentricity in a similar way as flow harmonics, since flow harmonics stemed from the initial stage are expected to inherit mostly the $p_{T}$ dependences of the initial partonic anisotropies~\cite{Kolb200469}. Recent study suggests that initial hard partons play an important role in the final harmonic flow formation~\cite{Schulc2014}. $p_T$ tomographic study of the initial eccentricity is of great importance to check the anisotropy generation and afterburner development. In AMPT model, initial partons decomposed from excited strings and minijet partons carry all the phase space information providing ideal conditions for study of source properties~\cite{Nie100519}.

\begin{figure*}[htbp]
\centering
\includegraphics[scale=0.9]{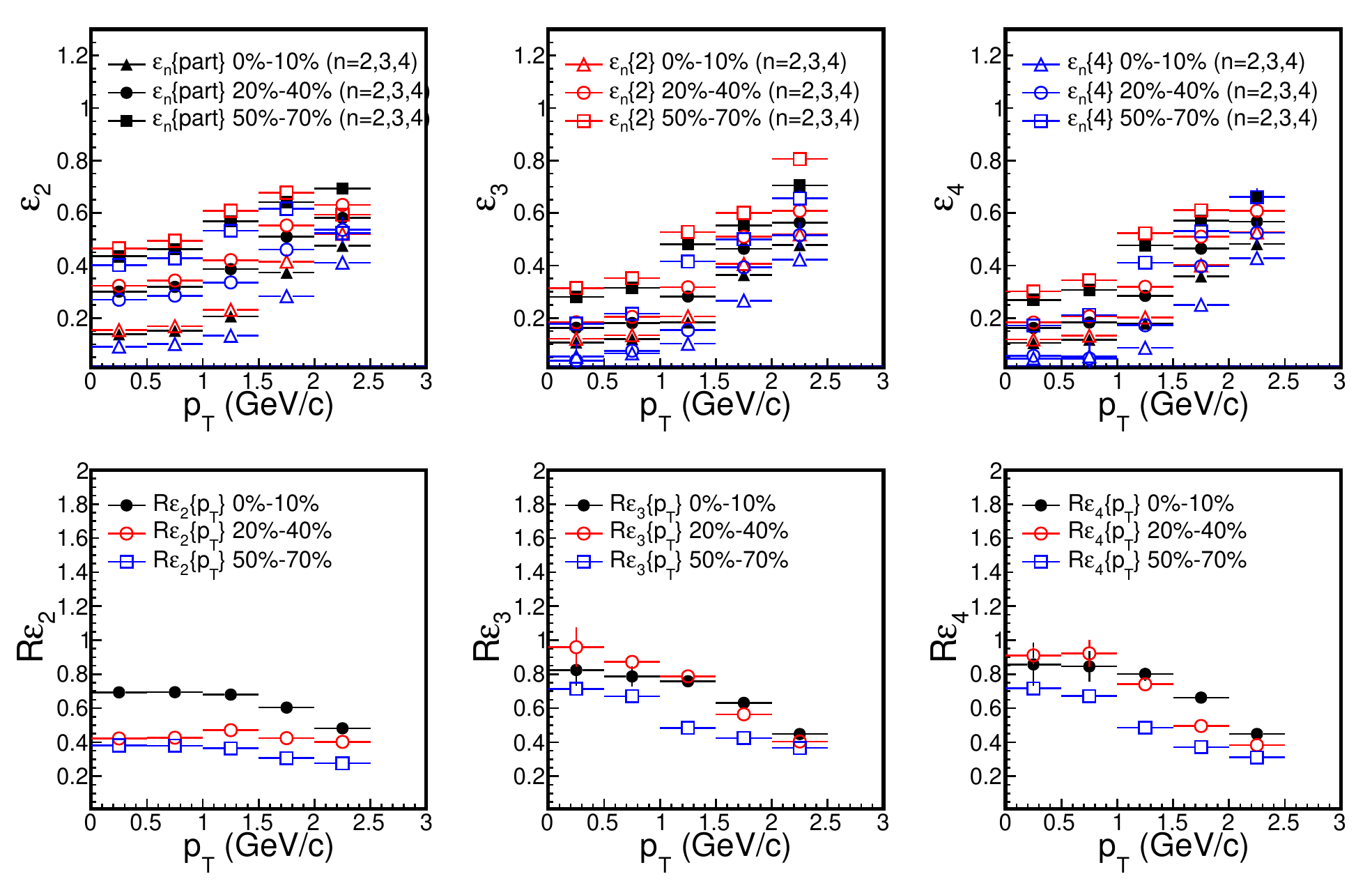}
\caption{(Color online) Eccentricity coefficients $\varepsilon_{n}$(n=2,3,4) defined with cumulant and participant method as a function of transverse momentum ($p_{T}$) for the AMPT initial condition. Upper panels: $\varepsilon_{n}$ defined by participant and regular cumulants as a functions of $p_{T}$. Lower panels: $\varepsilon_{n}$ (n$\geq$2) fluctuations as a functions of $p_{T}$, where $\varepsilon_{n}$ fluctuation is defined by Eq.(\ref{q14}). Results are shown for three different centrality classes in Au+Au collisions at $\sqrt{s_{NN}}$ = 200 GeV.}
\label{f5}
\end{figure*}

Figure~\ref{f5} shows initial partonic eccentricities and their fluctuations as a function of transverse momentum $p_{T}$ for three selected centrality classes. Similar $p_{T}$ dependences are for n=2, 3 and 4, i.e. initial eccentricity increases as a function of parton transverse momentum. A general inceasing trend can be observed for all the harmonics suggesting that higher $p_T$ partons contribute largely to the initial geometry anisotropy. The relative fluctuation of $\varepsilon_{n}$ ($R_{\varepsilon_{n}}\left\{p_{T}\right\}$) from regular cumulant definitions are seen to be smooth decreasing trend. $R_{\varepsilon_{2}}$ at low $p_T$ region is quite flatten which is quite similar to elliptic flow fluctuation. But, a deviation trend from flow fluctuation can be observed at higher $p_{T}$. Higher order $\varepsilon_{n}$(n$\geq$3) fluctuations show monotonic decreasing trend at high $p_T$. Direct comparison between initial eccentricity fluctuation and final flow fluctuation as a function of $p_{T}$ or $\eta$ may not be straightforward as partonic multi-scattering and final hadronic re-scattering after hadron freeze-out might bring in some substantial effects on the anisotropy development. Nevertheless, our results suggest that initial $\varepsilon_{n}$ fluctuation as a function of transverse momentum $p_{T}$ or pseudo-rapidity $\eta$ provides additional information of the source evolution. Further study on not only the pseudo-rapidity or transverse momentum dependence of $\varepsilon_{n}$ but also correlations between quantities at different transverse momentum or rapidity bins with AMPT model simulation will provide comprehensive understanding for the source anisotropy as motivated by studies~\cite{Khachatryan2015,Pang201552}.

\subsection{Harmonic flow response to the initial eccentricity in the AMPT model}

\begin{figure*}[htbp]
\centering
\includegraphics[scale=1.2]{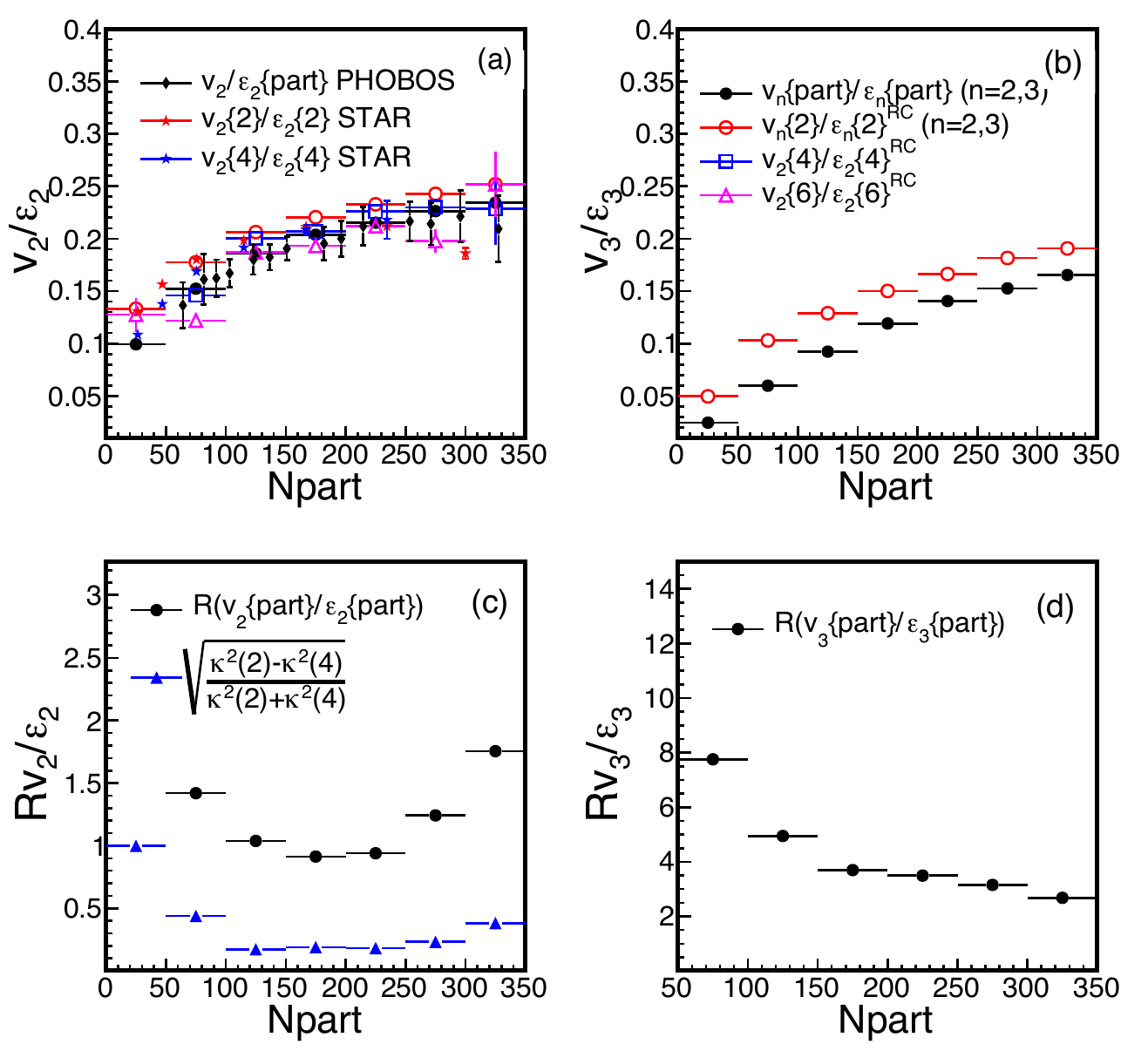}
\caption{(Color online) Conversion coefficient $v_{n}/\varepsilon_{n}$(n=2,3) and their fluctuation as a function of $N_{part}$. $v_{n}\left\{part\right\}/\varepsilon_{n}\left\{part\right\}$ from participant method and $v_{n}\left\{k\right\}/\varepsilon_{n}\left\{k\right\}$(k=2, 4, 6) from the cumulant method (Eq.(\ref{q7})). Upper Panels (a)-(b): $v_{n}/\varepsilon_{n}$ versus $N_{part}$ for different harmonic orders. The experimental results of $v_{n}\left\{2\right\}/\varepsilon_{n}\left\{2\right\}$ and $v_{n}\left\{4\right\}/\varepsilon_{n}\left\{4\right\}$ with flow measured in Q-cumulant method and eccentricity in regular cumulant method are shown for comparison. Lower Panels (c)-(d): Fluctuation of conversion coefficient $v_{n}/\varepsilon_{n}$(n=2,3) as a function of $N_{part}$, where $\kappa(2)$=$v_{2}\left\{2\right\}/\varepsilon_{2}\left\{2\right\}$ and $\kappa(4)$=$v_{2}\left\{4\right\}/\varepsilon_{2}\left\{4\right\}$.}
\label{f3}
\end{figure*}

In ideal hydrodynamics, a linear correlation is predicted between initial source geometric anisotropy and final flow of hadrons. In the past few years, impressive progresses have been made in studying flow response to the initial stage~\cite{Petersen2012,Song2011,Lacey2014112}. We understand that elliptic flow $v_2$ and triangular flow $v_3$ are driven mainly by the linear response to the initially produced fireball. For higher order harmonics, due to non-linear responses, the conversion of the initial geometry to the final flow becomes much more complicated which need to consider combinatorial contributions from different order of eccentricity harmionics as suggested by realistic simulation study~\cite{Gardim2012}. Taking the ratio $v_n/\varepsilon_n$ as the conversion coefficient from the initial eccentricity to the final flow, we further studied the ratio $v_{n}\left\{k\right\}/\varepsilon_{n}\left\{k\right\}$(k=4,6) with cumulant method and compared with results from participant method.  In Figure~\ref{f3}, we plot the conversion coefficient $v_{n}/\varepsilon_{n}$(n=2,3) as a function of number of participant nucleons $N_{part}$.  $v_{n}\left\{k\right\}/\varepsilon_{n}\left\{k\right\}$ (k$\geq$2) are based on cumulant definition Eq.(\ref{q7}). Experimental measurements of $v_{2}\left\{2\right\}/\varepsilon_{2}\left\{2\right\}$ and $v_{2}\left\{4\right\}/\varepsilon_{2}\left\{4\right\}$ with elliptic flow $v_{2}$ measured with Q-cumulant method where $\varepsilon_{2}$ with regular cumulant method based on MC-Glauber model are shown for comparison.

In the similar way as in experimental measurements, when $v_{2}\left\{k\right\}$(k=2,4) scaled with cumulant $\varepsilon_{2}\left\{k\right\}$(k=2,4), AMPT well reproduces the experimental results. The conversion coefficient from participant definition $v_{2}\left\{part\right\}/\varepsilon_{2}\left\{part\right\}$ follows similar trend as cumulant $v_{2}\left\{k\right\}/\varepsilon_{2}\left\{k\right\}$(k=2,4,6). The trend of $v_{n}/\varepsilon_{n}$ shows the hierarchy that $v_{2}\left\{2\right\}/\varepsilon_{2}\left\{2\right\}$ is systematically higher than higher order cumulant $v_{2}\left\{k\right\}/\varepsilon_{2}\left\{k\right\}$(k=4,6) over the whole centrality region. Further specific study of the conversion coefficient of the initial profile with considering linear and non-linear hydrodynamic responses is expected to provide more qualitative descriptions~\cite{Jia201441,Gardim2012}.

\begin{figure*}[htbp]
\centering
\includegraphics[scale=1.2]{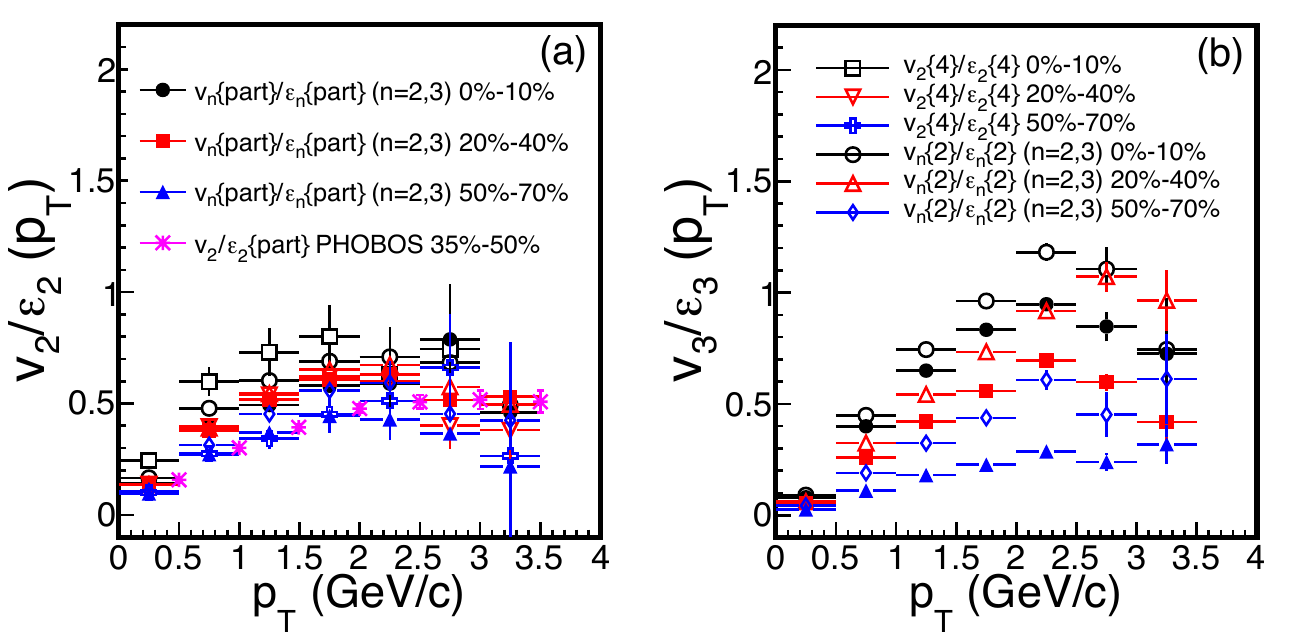}
\caption{(Color online) Conversion coefficients $v_{n}/\varepsilon_{n}$(n=2,3)  as a function of transverse momentum $p_{T}$. $v_{n}/\varepsilon_{n}$ from participant method and regular cumulant method are studied.}
\label{f6}
\end{figure*}

\begin{figure*}[htbp]
\centering
\includegraphics[scale=1.2]{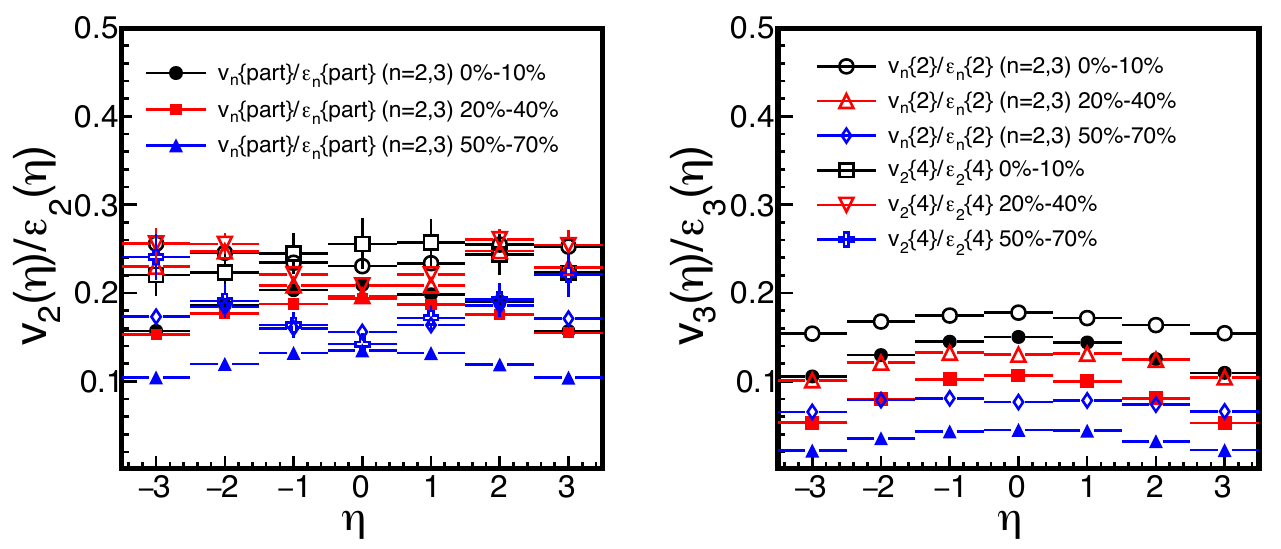}
\caption{(Color online) Conversion coefficients $v_{n}/\varepsilon_{n}$(n=2,3) as a function of pseudo-rapidity $\eta$. $v_{n}/\varepsilon_{n}$ from traditional participant method and regular cumulant method (RC) are studied.}
\label{f7}
\end{figure*}

Recent studies suggest that hard probes like jet are prospective for tomographic study of initial source profile and harmonic fluctuations in the initial states~\cite{Zhang201489,Zhang2012713,Nie2014}. Motivated by this idea, we study final hadron flow responses to the initial parton eccentricity as a function of transverse momentum $p_{T}$. Figure~\ref{f6} shows $p_{T}$ dependence of the coefficient $v_{n}/\varepsilon_{n}$ from both cumulant method and participant method. We can see that $v_{n}(p_{T})/\varepsilon_{n}$ generally show an increasing trend as a function of $p_{T}$. For both cumulant and participant $v_{n}/\varepsilon_{n}$, one can see that the conversion efficiency tends to be larger at higher $p_{T}$. Centrality dependence of the $v_{n}(p_{T})/\varepsilon_{n}$ is presented by investigating three centrality classes from central to peripheral collision. More simulation data is needed to extend to even higher $p_{T}$ region to study hard jet response. In additional to the study of $p_{T}$ dependence, the pseudo-rapidity $\eta$ dependence of $v_{n}/\varepsilon_{n}$ (n=2,3) are also studied by looking into the ratio of $v_{n}(\eta)$ to $\varepsilon_{n}(\eta)$ at corresponding rapidity region. The flow response to the initial eccentricity in the longitudinal direction are studied. Results of $v_{n}(\eta)/\varepsilon_{n}(\eta)$ are shown in Fig.\ref{f7} with symmetric shape observed.  Both $v_{n}(\eta)/\varepsilon_{n}(\eta)$(n=2,3) from both participant definition and regular cumulant definition are found quite similar to the distribution of corresponding $v_{n}(\eta)$ which shows slight $\eta$ dependence. Cumulant $v_{n}(\eta)/\varepsilon_{n}(\eta)$(n=2,3) shows weaker $\eta$ dependence in comparison with participant $v_{n}(\eta)/\varepsilon_{n}(\eta)$(n=2,3) which suggests the proportionality between $\varepsilon_{n}$ at fixed spatial rapidity and $v_{n}$ at fixed pseudorapidity changes little in the longitudinal direction.

\subsection{ Partonic effect on the eccentricity and eccentricity fluctuation }

\begin{figure*}[htbp]
\centering
\includegraphics[scale=0.9]{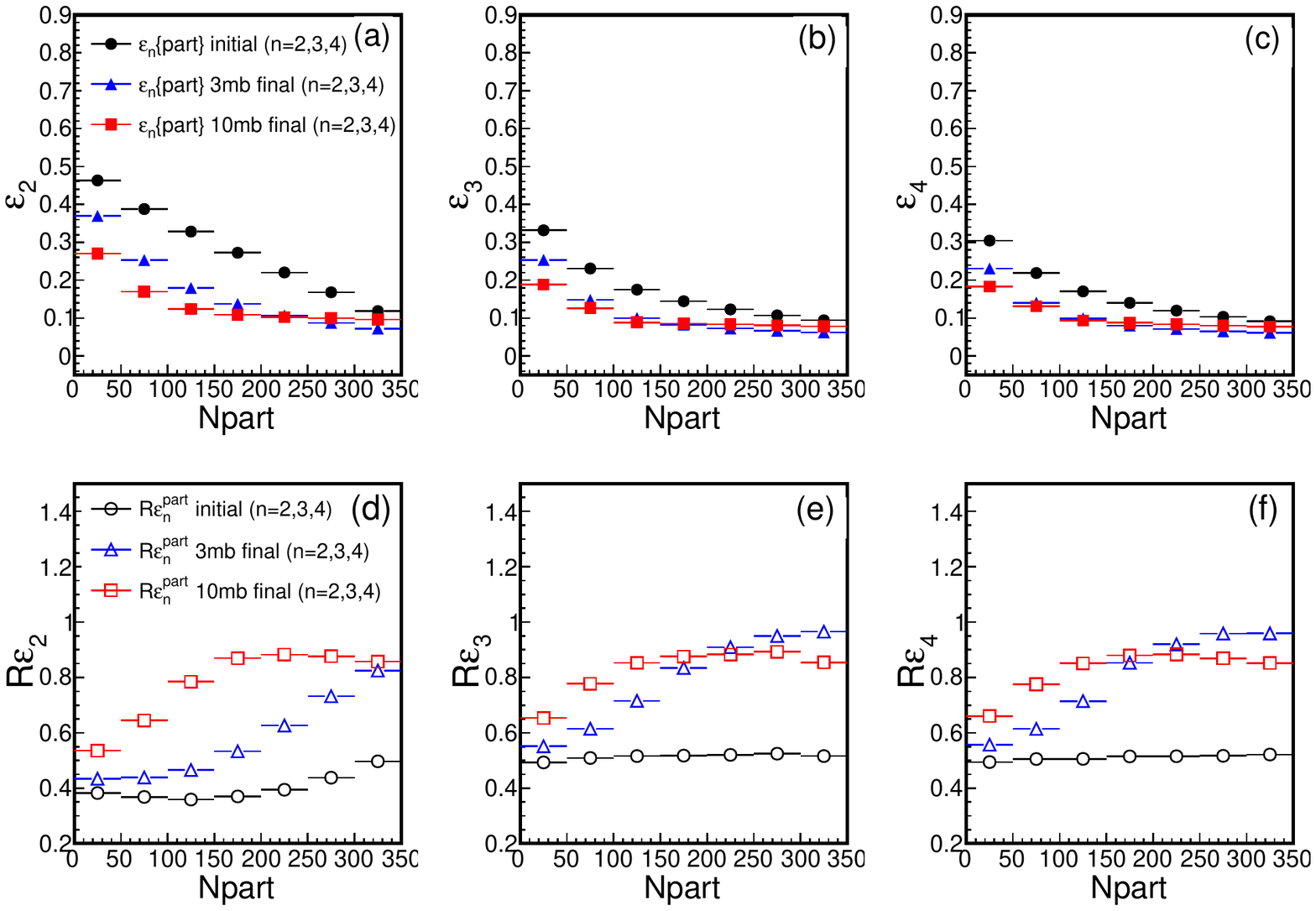}
\caption{(Color online) Eccentricity and fluctuation as a function of mean value of participant nucleons $N_{part}$ in AMPT model for Au+Au collision at 200 GeV. Upper panels: $\varepsilon_{n}\left\{part\right\}$ before (denote as 'initial') and after (denote as 'final') partonic scatterings with two parton cross section settings of 3mb and 10mb. Lower panels: $\varepsilon_{n}\left\{part\right\}$ fluctuation $R_{\varepsilon_{n}^{part}}$ before and after partonic scatterings.}
\label{f8}
\end{figure*}

The parton scattering cross section in the AMPT model has shown considerable influence on the magnitude of the flow coefficients~\cite{Chen2004}. It is important to investigate the effect on eccentricity and eccentricity fluctuation in the partonic stage since it may shed light on the evolution dynamics of the source in heavy-ion collision.  Figure~\ref{f8} upper pannel shows the source participant eccentricity before (denote as initial) and after (denote as final) partonic scatterings for different orders of harmonic as a function of mean value of participant nucleons for partons in Au+Au collisions from the AMPT simulations. Parton scattering cross sections were selected as 3 mb and 10 mb.  It indicates that partonic scattering significantly reduces eccentricity commonly for all the order of harmonics. We find for non-central collsions the larger partonic cross section, the smaller final $\varepsilon_{n}$ after partonic scattering but for central collisions partonic scattering cross section makes little effect on the source eccentricity $\varepsilon_{n}$.

Relative fluctuation of participant $\varepsilon_{n}\left\{part\right\}$ - $R_{\varepsilon_{n}^{part}}$(n$\geq$2) are shown in the lower pannels of Figure~\ref{f8}, where we give the comparison of fluctuation $R_{\varepsilon_{n}^{part}}$ before and after partonic scatterings with two different partonic cross sections. Partonic scattering dramatically increases fluctuation of $\varepsilon_{n}\left\{part\right\}$ for different order of harmonics.  Experimental measurements of higher order flow fluctuations with cumulant method will be prospective for quantitatively understanding of development of anisotropy fluctuation from initial partonic stage to the final hadronic stage.

\section{Summary}

In summary, in the framework of a multi-phase transport model (AMPT), initial partonic eccentricity and eccentricity fluctuations are studied up to sixth order of harmonic by means of traditional participant method and multi-particle cumulant method in Au+Au collisions at center-of-mass energy of 200 GeV. Eccentricities $\varepsilon_{n}$ and fluctuations $R_{\varepsilon_{n}}$ defined with participant method and regular cumulant method are studied and compared systematically.  Eccentricity fluctuation shows a similar picture as flow fluctuation which confirms the close relationship between initial eccentricity harmonics and final flow harmonics. Flow responses are investigated by the ratio $v_{n}/\varepsilon_{n}$ as a function of number of participant nucleons ($N_{part}$), transverse momentum ($p_{T}$) and pseudo-rapidity($\eta$) for a tomographic study of the conversion properties. Relative fluctuations of $\varepsilon_{n}$ defined by cumulants as a function of transverse momentum and pseudo-rapidity are also studied specifically for the transverse and longitudinal features of the created source. $\varepsilon_{n}$ fluctuation versus $p_T$ and $\eta$ show similar trends as corresponding flow harmonic and flow fluctuation measured experimentally. Higher harmonic eccentricity fluctuation studies are expected to give further constraint to higher order harmonic flow studies.

Similar to anisotropic flow measurements which have been proved to be sensitive to the shape and shape fluctuation of the initial overlap zone, direct measurements of eccentricity fluctuations could lead to a better understanding of the initial source conditions. Through the comparison of the AMPT model simulation results with experimental measurements, we found that ellipticity and triangularity as well as higher harmonic initial anisotropies show similar behaviours as final flow harmonics both in the transverse and longitudinal directions. As event-by-event fluctuations are crucial to the current understanding of relativistic heavy-ion collision, the study on physics origin of how fluctuations of flow harmonics stems from the early stage of collision will be of great importance. The AMPT model simulations provide a promising way of studying initial partonic state. Future experimental study of anisotropic flow harmonics with extended $p_T$ region and $\eta$ region can provide further constrain on the initial source profile. We also expect studies on the initial fluctuations in smaller system like p+Au, d+Au or $^3$He+Au especially fluctuation properties in the longitudinal direction can bring complementary information of the source evolution mechanisms.

\section*{Acknowledgements}
This work is supported by the Major State Basic Research Development Program in China under Grant No. 2014CB845400, the National Natural Science Foundation of China under Grants Nos. 11421505, 11220101005, 11522547 and 11375251, the Youth Innovation Promotion Association of CAS under Grant No. 2013175.



\begin{thebibliography}{99}

\expandafter\ifx\csname natexlab\endcsname\relax\def\natexlab#1{#1}\fi
\expandafter\ifx\csname bibnamefont\endcsname\relax
  \def\bibnamefont#1{#1}\fi
\expandafter\ifx\csname bibfnamefont\endcsname\relax
  \def\bibfnamefont#1{#1}\fi
\expandafter\ifx\csname citenamefont\endcsname\relax
  \def\citenamefont#1{#1}\fi
\expandafter\ifx\csname url\endcsname\relax
  \def\url#1{\texttt{#1}}\fi
\expandafter\ifx\csname urlprefix\endcsname\relax\def\urlprefix{URL }\fi
\providecommand{\bibinfo}[2]{#2}
\providecommand{\eprint}[2][]{\url{#2}}

\bibitem[{\citenamefont{Ollitrault}(1992)}]{Ollitrault1992}
\bibinfo{author}{\bibfnamefont{J.-Y.} \bibnamefont{Ollitrault}},
  \bibinfo{journal}{Phys. Rev. D} \textbf{\bibinfo{volume}{46}},
  \bibinfo{pages}{229} (\bibinfo{year}{1992}),
  \urlprefix\url{http://link.aps.org/doi/10.1103/PhysRevD.46.229}.

\bibitem[{\citenamefont{Voloshin
  et~al.}(2008{\natexlab{a}})\citenamefont{Voloshin, Poskanzer, and
  Snellings}}]{Voloshin2008}
\bibinfo{author}{\bibfnamefont{S.~A.} \bibnamefont{Voloshin}},
  \bibinfo{author}{\bibfnamefont{A.~M.} \bibnamefont{Poskanzer}},
  \bibnamefont{and} \bibinfo{author}{\bibfnamefont{R.}~\bibnamefont{Snellings}}
  (\bibinfo{year}{2008}{\natexlab{a}}), \eprint{0809.2949}.

\bibitem[{\citenamefont{Kolb et~al.}(1999)\citenamefont{Kolb, Sollfrank, and
  Heinz}}]{Kolb1999}
\bibinfo{author}{\bibfnamefont{P.~F.} \bibnamefont{Kolb}},
  \bibinfo{author}{\bibfnamefont{J.}~\bibnamefont{Sollfrank}},
  \bibnamefont{and} \bibinfo{author}{\bibfnamefont{U.~W.} \bibnamefont{Heinz}},
  \bibinfo{journal}{Phys. Lett.} \textbf{\bibinfo{volume}{B459}},
  \bibinfo{pages}{667} (\bibinfo{year}{1999}).

\bibitem[{\citenamefont{Ackermann et~al.}(2001)}]{Ackermann2001}
\bibinfo{author}{\bibfnamefont{K.~H.} \bibnamefont{Ackermann}}
  \bibnamefont{et~al.} (\bibinfo{collaboration}{STAR Collaboration}),
  \bibinfo{journal}{Phys. Rev. Lett.} \textbf{\bibinfo{volume}{86}},
  \bibinfo{pages}{402} (\bibinfo{year}{2001}),
  \urlprefix\url{http://link.aps.org/doi/10.1103/PhysRevLett.86.402}.

\bibitem[{\citenamefont{Teaney et~al.}(2001)\citenamefont{Teaney, Lauret, and
  Shuryak}}]{Teaney2001}
\bibinfo{author}{\bibfnamefont{D.}~\bibnamefont{Teaney}},
  \bibinfo{author}{\bibfnamefont{J.}~\bibnamefont{Lauret}}, \bibnamefont{and}
  \bibinfo{author}{\bibfnamefont{E.~V.} \bibnamefont{Shuryak}},
  \bibinfo{journal}{Phys. Rev. Lett.} \textbf{\bibinfo{volume}{86}},
  \bibinfo{pages}{4783} (\bibinfo{year}{2001}),
  \urlprefix\url{http://link.aps.org/doi/10.1103/PhysRevLett.86.4783}.

\bibitem[{\citenamefont{Romatschke and Romatschke}(2007)}]{Romatschke2007}
\bibinfo{author}{\bibfnamefont{P.}~\bibnamefont{Romatschke}} \bibnamefont{and}
  \bibinfo{author}{\bibfnamefont{U.}~\bibnamefont{Romatschke}},
  \bibinfo{journal}{Phys. Rev. Lett.} \textbf{\bibinfo{volume}{99}},
  \bibinfo{pages}{172301} (\bibinfo{year}{2007}),
  \urlprefix\url{http://link.aps.org/doi/10.1103/PhysRevLett.99.172301}.

\bibitem[{\citenamefont{Song et~al.}(2011)\citenamefont{Song, Bass, Heinz,
  Hirano, and Shen}}]{Song2011}
\bibinfo{author}{\bibfnamefont{H.}~\bibnamefont{Song}},
  \bibinfo{author}{\bibfnamefont{S.~A.} \bibnamefont{Bass}},
  \bibinfo{author}{\bibfnamefont{U.}~\bibnamefont{Heinz}},
  \bibinfo{author}{\bibfnamefont{T.}~\bibnamefont{Hirano}}, \bibnamefont{and}
  \bibinfo{author}{\bibfnamefont{C.}~\bibnamefont{Shen}},
  \bibinfo{journal}{Phys. Rev. Lett.} \textbf{\bibinfo{volume}{106}},
  \bibinfo{pages}{192301} (\bibinfo{year}{2011}),
  \urlprefix\url{http://link.aps.org/doi/10.1103/PhysRevLett.106.192301}.

\bibitem{Ko}C. M. Ko  et al., Nuclear Science and Techniques {\bf 24},  050525 (2013).

\bibitem[{\citenamefont{Heinz}(2005)}]{Ulrich2005}
\bibinfo{author}{\bibfnamefont{U.}~\bibnamefont{Heinz}},
  \bibinfo{journal}{Journal of Physics G: Nuclear and Particle Physics}
  \textbf{\bibinfo{volume}{31}}, \bibinfo{pages}{S717} (\bibinfo{year}{2005}),
  \urlprefix\url{http://stacks.iop.org/0954-3899/31/i=6/a=012}.

\bibitem[{\citenamefont{Gale et~al.}(2013)\citenamefont{Gale, Jeon, and
  Schenke}}]{Gale2013}
\bibinfo{author}{\bibfnamefont{C.}~\bibnamefont{Gale}},
  \bibinfo{author}{\bibfnamefont{S.}~\bibnamefont{Jeon}}, \bibnamefont{and}
  \bibinfo{author}{\bibfnamefont{B.}~\bibnamefont{Schenke}},
  \bibinfo{journal}{Int. J. Mod. Phys.} \textbf{\bibinfo{volume}{A28}},
  \bibinfo{pages}{1340011} (\bibinfo{year}{2013}).

\bibitem[{\citenamefont{Qiu and Heinz}(2012)}]{Qiu20111441}
\bibinfo{author}{\bibfnamefont{Z.}~\bibnamefont{Qiu}} \bibnamefont{and}
  \bibinfo{author}{\bibfnamefont{U.}~\bibnamefont{Heinz}},
  \bibinfo{journal}{AIP Conf. Proc.} \textbf{\bibinfo{volume}{1441}},
  \bibinfo{pages}{774} (\bibinfo{year}{2012}), \eprint{1108.1714}.

\bibitem[{\citenamefont{Adams et~al.}(2004)\citenamefont{Adams, Adler
  et~al.}}]{STAR2004flow}
\bibinfo{author}{\bibfnamefont{J.}~\bibnamefont{Adams}},
  \bibinfo{author}{\bibfnamefont{C.}~\bibnamefont{Adler}}, \bibnamefont{et~al.}
  (\bibinfo{collaboration}{STAR Collaboration}), \bibinfo{journal}{Phys. Rev.
  Lett.} \textbf{\bibinfo{volume}{92}}, \bibinfo{pages}{062301}
  (\bibinfo{year}{2004}),
  \urlprefix\url{http://link.aps.org/doi/10.1103/PhysRevLett.92.062301}.

\bibitem[{\citenamefont{Adamczyk et~al.}(2013)}]{STAR2013flow}
\bibinfo{author}{\bibfnamefont{L.}~\bibnamefont{Adamczyk}} \bibnamefont{et~al.}
  (\bibinfo{collaboration}{STAR Collaboration}), \bibinfo{journal}{Phys. Rev.}
  \textbf{\bibinfo{volume}{C88}}, \bibinfo{pages}{014904}
  (\bibinfo{year}{2013}).

\bibitem[{\citenamefont{Adare et~al.}(2011)}]{PHENIX2011flow}
\bibinfo{author}{\bibfnamefont{A.}~\bibnamefont{Adare}} \bibnamefont{et~al.}
  (\bibinfo{collaboration}{PHENIX Collaboration}), \bibinfo{journal}{Phys. Rev.
  Lett.} \textbf{\bibinfo{volume}{107}}, \bibinfo{pages}{252301}
  (\bibinfo{year}{2011}).

\bibitem[{\citenamefont{Chen et~al.}(2004)\citenamefont{Chen, Ko, and
  Lin}}]{Chen2004}
\bibinfo{author}{\bibfnamefont{L.-W.} \bibnamefont{Chen}},
  \bibinfo{author}{\bibfnamefont{C.~M.} \bibnamefont{Ko}}, \bibnamefont{and}
  \bibinfo{author}{\bibfnamefont{Z.-W.} \bibnamefont{Lin}},
  \bibinfo{journal}{Phys. Rev. C} \textbf{\bibinfo{volume}{69}},
  \bibinfo{pages}{031901} (\bibinfo{year}{2004}),
  \urlprefix\url{http://link.aps.org/doi/10.1103/PhysRevC.69.031901}.

\bibitem[{\citenamefont{Han et~al.}(2011)\citenamefont{Han, Ma, Ma, Cai, Chen,
  Zhang, and Zhong}}]{Han2011}
\bibinfo{author}{\bibfnamefont{L.~X.} \bibnamefont{Han}},
  \bibinfo{author}{\bibfnamefont{G.~L.} \bibnamefont{Ma}},
  \bibinfo{author}{\bibfnamefont{Y.~G.} \bibnamefont{Ma}},
  \bibinfo{author}{\bibfnamefont{X.~Z.} \bibnamefont{Cai}},
  \bibinfo{author}{\bibfnamefont{J.~H.} \bibnamefont{Chen}},
  \bibinfo{author}{\bibfnamefont{S.}~\bibnamefont{Zhang}}, \bibnamefont{and}
  \bibinfo{author}{\bibfnamefont{C.}~\bibnamefont{Zhong}},
  \bibinfo{journal}{Phys. Rev. C} \textbf{\bibinfo{volume}{84}},
  \bibinfo{pages}{064907} (\bibinfo{year}{2011}),
  \urlprefix\url{http://link.aps.org/doi/10.1103/PhysRevC.84.064907}.
  
  \bibitem{Phenix2}A. Adare et al. (PHENIX Collaboration), Phys. Rev. C {\bf 93}, 051902(R) (2016).

\bibitem[{\citenamefont{Alver and Roland}(2010)}]{Alver201081}
\bibinfo{author}{\bibfnamefont{B.}~\bibnamefont{Alver}} \bibnamefont{and}
  \bibinfo{author}{\bibfnamefont{G.}~\bibnamefont{Roland}},
  \bibinfo{journal}{Phys. Rev. C} \textbf{\bibinfo{volume}{81}},
  \bibinfo{pages}{054905} (\bibinfo{year}{2010}),
  \urlprefix\url{http://link.aps.org/doi/10.1103/PhysRevC.81.054905}.

\bibitem[{\citenamefont{Schenke et~al.}(2011)\citenamefont{Schenke, Jeon, and
  Gale}}]{Schenke2011}
\bibinfo{author}{\bibfnamefont{B.}~\bibnamefont{Schenke}},
  \bibinfo{author}{\bibfnamefont{S.}~\bibnamefont{Jeon}}, \bibnamefont{and}
  \bibinfo{author}{\bibfnamefont{C.}~\bibnamefont{Gale}},
  \bibinfo{journal}{Phys. Rev. Lett.} \textbf{\bibinfo{volume}{106}},
  \bibinfo{pages}{042301} (\bibinfo{year}{2011}),
  \urlprefix\url{http://link.aps.org/doi/10.1103/PhysRevLett.106.042301}.

\bibitem[{\citenamefont{Schenke et~al.}(2012)\citenamefont{Schenke, Jeon, and
  Gale}}]{Schenke2012}
\bibinfo{author}{\bibfnamefont{B.}~\bibnamefont{Schenke}},
  \bibinfo{author}{\bibfnamefont{S.}~\bibnamefont{Jeon}}, \bibnamefont{and}
  \bibinfo{author}{\bibfnamefont{C.}~\bibnamefont{Gale}},
  \bibinfo{journal}{Phys. Rev. C} \textbf{\bibinfo{volume}{85}},
  \bibinfo{pages}{024901} (\bibinfo{year}{2012}),
  \urlprefix\url{http://link.aps.org/doi/10.1103/PhysRevC.85.024901}.

\bibitem[{\citenamefont{Andrade et~al.}(2006)\citenamefont{Andrade, Grassi,
  Hama, Kodama, and Socolowski}}]{Andrade2006}
\bibinfo{author}{\bibfnamefont{R.}~\bibnamefont{Andrade}},
  \bibinfo{author}{\bibfnamefont{F.}~\bibnamefont{Grassi}},
  \bibinfo{author}{\bibfnamefont{Y.}~\bibnamefont{Hama}},
  \bibinfo{author}{\bibfnamefont{T.}~\bibnamefont{Kodama}}, \bibnamefont{and}
  \bibinfo{author}{\bibfnamefont{O.}~\bibnamefont{Socolowski}},
  \bibinfo{journal}{Phys. Rev. Lett.} \textbf{\bibinfo{volume}{97}},
  \bibinfo{pages}{202302} (\bibinfo{year}{2006}),
  \urlprefix\url{http://link.aps.org/doi/10.1103/PhysRevLett.97.202302}.

\bibitem[{\citenamefont{Petersen et~al.}(2010)\citenamefont{Petersen, Qin,
  Bass, and M\"uller}}]{Petersen2010}
\bibinfo{author}{\bibfnamefont{H.}~\bibnamefont{Petersen}},
  \bibinfo{author}{\bibfnamefont{G.-Y.} \bibnamefont{Qin}},
  \bibinfo{author}{\bibfnamefont{S.~A.} \bibnamefont{Bass}}, \bibnamefont{and}
  \bibinfo{author}{\bibfnamefont{B.}~\bibnamefont{M\"uller}},
  \bibinfo{journal}{Phys. Rev. C} \textbf{\bibinfo{volume}{82}},
  \bibinfo{pages}{041901} (\bibinfo{year}{2010}),
  \urlprefix\url{http://link.aps.org/doi/10.1103/PhysRevC.82.041901}.

\bibitem[{\citenamefont{Ma et~al.}(2014)\citenamefont{Ma, Ma, and Ma}}]{Ma2014}
\bibinfo{author}{\bibfnamefont{L.}~\bibnamefont{Ma}},
  \bibinfo{author}{\bibfnamefont{G.~L.} \bibnamefont{Ma}}, \bibnamefont{and}
  \bibinfo{author}{\bibfnamefont{Y.~G.} \bibnamefont{Ma}},
  \bibinfo{journal}{Phys. Rev. C} \textbf{\bibinfo{volume}{89}},
  \bibinfo{pages}{044907} (\bibinfo{year}{2014}),
  \urlprefix\url{http://link.aps.org/doi/10.1103/PhysRevC.89.044907}.

\bibitem[{\citenamefont{Sorensen}(2007)}]{Sorensen2007}
\bibinfo{author}{\bibfnamefont{P.}~\bibnamefont{Sorensen}},
  \bibinfo{journal}{Journal of Physics G: Nuclear and Particle Physics}
  \textbf{\bibinfo{volume}{34}}, \bibinfo{pages}{S897} (\bibinfo{year}{2007}),
  \urlprefix\url{http://stacks.iop.org/0954-3899/34/i=8/a=S121}.

\bibitem[{\citenamefont{Alver et~al.}(2007{\natexlab{a}})}]{Alver200734}
\bibinfo{author}{\bibfnamefont{B.}~\bibnamefont{Alver}} \bibnamefont{et~al.}
  (\bibinfo{collaboration}{PHOBOS Collaboration}), \bibinfo{journal}{J. Phys.}
  \textbf{\bibinfo{volume}{G34}}, \bibinfo{pages}{S907}
  (\bibinfo{year}{2007}{\natexlab{a}}).

\bibitem[{\citenamefont{Alver et~al.}(2010{\natexlab{a}})\citenamefont{Alver,
  Back et~al.}}]{Alver2010104}
\bibinfo{author}{\bibfnamefont{B.}~\bibnamefont{Alver}},
  \bibinfo{author}{\bibfnamefont{B.~B.} \bibnamefont{Back}},
  \bibnamefont{et~al.}, \bibinfo{journal}{Phys. Rev. Lett.}
  \textbf{\bibinfo{volume}{104}}, \bibinfo{pages}{142301}
  (\bibinfo{year}{2010}{\natexlab{a}}),
  \urlprefix\url{http://link.aps.org/doi/10.1103/PhysRevLett.104.142301}.

\bibitem[{\citenamefont{Agakishiev et~al.}(2012)}]{Agakishiev2011}
\bibinfo{author}{\bibfnamefont{G.}~\bibnamefont{Agakishiev}}
  \bibnamefont{et~al.} (\bibinfo{collaboration}{STAR Collaboration}),
  \bibinfo{journal}{Phys. Rev.} \textbf{\bibinfo{volume}{C86}},
  \bibinfo{pages}{014904} (\bibinfo{year}{2012}).

\bibitem[{\citenamefont{Holopainen et~al.}(2011)\citenamefont{Holopainen,
  Niemi, and Eskola}}]{Holopainen2010}
\bibinfo{author}{\bibfnamefont{H.}~\bibnamefont{Holopainen}},
  \bibinfo{author}{\bibfnamefont{H.}~\bibnamefont{Niemi}}, \bibnamefont{and}
  \bibinfo{author}{\bibfnamefont{K.~J.} \bibnamefont{Eskola}},
  \bibinfo{journal}{Phys. Rev.} \textbf{\bibinfo{volume}{C83}},
  \bibinfo{pages}{034901} (\bibinfo{year}{2011}), \eprint{1007.0368}.

\bibitem[{\citenamefont{Alver et~al.}(2010{\natexlab{b}})\citenamefont{Alver,
  Gombeaud, Luzum, and Ollitrault}}]{Alver201082}
\bibinfo{author}{\bibfnamefont{B.~H.} \bibnamefont{Alver}},
  \bibinfo{author}{\bibfnamefont{C.}~\bibnamefont{Gombeaud}},
  \bibinfo{author}{\bibfnamefont{M.}~\bibnamefont{Luzum}}, \bibnamefont{and}
  \bibinfo{author}{\bibfnamefont{J.-Y.} \bibnamefont{Ollitrault}},
  \bibinfo{journal}{Phys. Rev.} \textbf{\bibinfo{volume}{C82}},
  \bibinfo{pages}{034913} (\bibinfo{year}{2010}{\natexlab{b}}).

\bibitem[{\citenamefont{Miller and Snellings}(2003)}]{Miller2003}
\bibinfo{author}{\bibfnamefont{M.}~\bibnamefont{Miller}} \bibnamefont{and}
  \bibinfo{author}{\bibfnamefont{R.}~\bibnamefont{Snellings}}
  (\bibinfo{year}{2003}), \eprint{nucl-ex/0312008}.

\bibitem[{\citenamefont{Derradi~de Souza et~al.}(2012)\citenamefont{Derradi~de
  Souza, Takahashi, Kodama, and Sorensen}}]{DerradideSouza2011}
\bibinfo{author}{\bibfnamefont{R.}~\bibnamefont{Derradi~de Souza}},
  \bibinfo{author}{\bibfnamefont{J.}~\bibnamefont{Takahashi}},
  \bibinfo{author}{\bibfnamefont{T.}~\bibnamefont{Kodama}}, \bibnamefont{and}
  \bibinfo{author}{\bibfnamefont{P.}~\bibnamefont{Sorensen}},
  \bibinfo{journal}{Phys. Rev.} \textbf{\bibinfo{volume}{C85}},
  \bibinfo{pages}{054909} (\bibinfo{year}{2012}), \eprint{1110.5698}.

\bibitem[{\citenamefont{Ma and Wang}(2011)}]{Ma2011106}
\bibinfo{author}{\bibfnamefont{G.-L.} \bibnamefont{Ma}} \bibnamefont{and}
  \bibinfo{author}{\bibfnamefont{X.-N.} \bibnamefont{Wang}},
  \bibinfo{journal}{Phys. Rev. Lett.} \textbf{\bibinfo{volume}{106}},
  \bibinfo{pages}{162301} (\bibinfo{year}{2011}),
  \urlprefix\url{http://link.aps.org/doi/10.1103/PhysRevLett.106.162301}.

\bibitem[{\citenamefont{Wang et~al.}(2013)\citenamefont{Wang, Ma, Zhang, Fang,
  Han, and Shen}}]{Wang2014}
\bibinfo{author}{\bibfnamefont{J.}~\bibnamefont{Wang}},
  \bibinfo{author}{\bibfnamefont{Y.}~\bibnamefont{Ma}},
  \bibinfo{author}{\bibfnamefont{G.}~\bibnamefont{Zhang}},
  \bibinfo{author}{\bibfnamefont{D.}~\bibnamefont{Fang}},
  \bibinfo{author}{\bibfnamefont{L.}~\bibnamefont{Han}}, \bibnamefont{and}
  \bibinfo{author}{\bibfnamefont{W.}~\bibnamefont{Shen}},
  \bibinfo{journal}{Nuclear Science and Techniques}
  \textbf{\bibinfo{volume}{24}}, \bibinfo{pages}{30501} (\bibinfo{year}{2013}),
  \urlprefix\url{http://www.j.sinap.ac.cn/nst/EN/abstract/article_13.shtml}.
  
\bibitem{WangJ2}  J. Wang, Y. G. Ma,  G. Q. Zhang, and W. Q. Shen, Phys. Rev. C {\bf 90}, 054601 (2014).
  \urlprefix\url{http://dx.doi.org/10.1103/PhysRevC.90.054601}


\bibitem[{\citenamefont{Alver et~al.}(2007{\natexlab{b}})\citenamefont{Alver,
  Back et~al.}}]{Alver200798}
\bibinfo{author}{\bibfnamefont{B.}~\bibnamefont{Alver}},
  \bibinfo{author}{\bibfnamefont{B.~B.} \bibnamefont{Back}},
  \bibnamefont{et~al.}, \bibinfo{journal}{Phys. Rev. Lett.}
  \textbf{\bibinfo{volume}{98}}, \bibinfo{pages}{242302}
  (\bibinfo{year}{2007}{\natexlab{b}}),
  \urlprefix\url{http://link.aps.org/doi/10.1103/PhysRevLett.98.242302}.

\bibitem[{\citenamefont{Drescher and Nara}(2007)}]{Drescher2007}
\bibinfo{author}{\bibfnamefont{H.-J.} \bibnamefont{Drescher}} \bibnamefont{and}
  \bibinfo{author}{\bibfnamefont{Y.}~\bibnamefont{Nara}},
  \bibinfo{journal}{Phys. Rev. C} \textbf{\bibinfo{volume}{76}},
  \bibinfo{pages}{041903} (\bibinfo{year}{2007}),
  \urlprefix\url{http://link.aps.org/doi/10.1103/PhysRevC.76.041903}.

\bibitem[{\citenamefont{Broniowski et~al.}(2007)\citenamefont{Broniowski,
  Bo\ifmmode~\dot{z}\else \.{z}\fi{}ek, and Rybczy\ifmmode~\acute{n}\else
  \'{n}\fi{}ski}}]{Broniowski2007}
\bibinfo{author}{\bibfnamefont{W.}~\bibnamefont{Broniowski}},
  \bibinfo{author}{\bibfnamefont{P.}~\bibnamefont{Bo\ifmmode~\dot{z}\else
  \.{z}\fi{}ek}}, \bibnamefont{and}
  \bibinfo{author}{\bibfnamefont{M.}~\bibnamefont{Rybczy\ifmmode~\acute{n}\else
  \'{n}\fi{}ski}}, \bibinfo{journal}{Phys. Rev. C}
  \textbf{\bibinfo{volume}{76}}, \bibinfo{pages}{054905}
  (\bibinfo{year}{2007}),
  \urlprefix\url{http://link.aps.org/doi/10.1103/PhysRevC.76.054905}.

\bibitem[{\citenamefont{Zhang et~al.}(2000)\citenamefont{Zhang, Ko, Li, and
  Lin}}]{Zhang2000}
\bibinfo{author}{\bibfnamefont{B.}~\bibnamefont{Zhang}},
  \bibinfo{author}{\bibfnamefont{C.~M.} \bibnamefont{Ko}},
  \bibinfo{author}{\bibfnamefont{B.-A.} \bibnamefont{Li}}, \bibnamefont{and}
  \bibinfo{author}{\bibfnamefont{Z.}~\bibnamefont{Lin}},
  \bibinfo{journal}{Phys. Rev. C} \textbf{\bibinfo{volume}{61}},
  \bibinfo{pages}{067901} (\bibinfo{year}{2000}),
  \urlprefix\url{http://link.aps.org/doi/10.1103/PhysRevC.61.067901}.

\bibitem[{\citenamefont{Wang and Gyulassy}(1991)}]{Wang1991}
\bibinfo{author}{\bibfnamefont{X.-N.} \bibnamefont{Wang}} \bibnamefont{and}
  \bibinfo{author}{\bibfnamefont{M.}~\bibnamefont{Gyulassy}},
  \bibinfo{journal}{Phys. Rev. D} \textbf{\bibinfo{volume}{44}},
  \bibinfo{pages}{3501} (\bibinfo{year}{1991}),
  \urlprefix\url{http://link.aps.org/doi/10.1103/PhysRevD.44.3501}.

\bibitem[{\citenamefont{Zhang}(1998)}]{Zhang1997}
\bibinfo{author}{\bibfnamefont{B.}~\bibnamefont{Zhang}},
  \bibinfo{journal}{Comput. Phys. Commun.} \textbf{\bibinfo{volume}{109}},
  \bibinfo{pages}{193} (\bibinfo{year}{1998}).

\bibitem[{\citenamefont{Lin et~al.}(2005)\citenamefont{Lin, Ko, Li, Zhang, and
  Pal}}]{Lin2005}
\bibinfo{author}{\bibfnamefont{Z.-W.} \bibnamefont{Lin}},
  \bibinfo{author}{\bibfnamefont{C.~M.} \bibnamefont{Ko}},
  \bibinfo{author}{\bibfnamefont{B.-A.} \bibnamefont{Li}},
  \bibinfo{author}{\bibfnamefont{B.}~\bibnamefont{Zhang}}, \bibnamefont{and}
  \bibinfo{author}{\bibfnamefont{S.}~\bibnamefont{Pal}},
  \bibinfo{journal}{Phys. Rev. C} \textbf{\bibinfo{volume}{72}},
  \bibinfo{pages}{064901} (\bibinfo{year}{2005}),
  \urlprefix\url{http://link.aps.org/doi/10.1103/PhysRevC.72.064901}.

\bibitem[{\citenamefont{Xu and Ko}(2011)}]{Xu2011}
\bibinfo{author}{\bibfnamefont{J.}~\bibnamefont{Xu}} \bibnamefont{and}
  \bibinfo{author}{\bibfnamefont{C.~M.} \bibnamefont{Ko}},
  \bibinfo{journal}{Phys. Rev. C} \textbf{\bibinfo{volume}{84}},
  \bibinfo{pages}{014903} (\bibinfo{year}{2011}),
  \urlprefix\url{http://link.aps.org/doi/10.1103/PhysRevC.84.014903}.



\bibitem{Ye}Y. J. Ye, J. H. Chen, Y. G. Ma,  S. Zhang,  and C. Zhong, Phys. Rev. C {\bf 93}, 044904 (2016).
 \urlprefix\url{http://link.aps.org/doi/10.1103/PhysRevC.93.044904}.


\bibitem{Xu}Yi-Fei Xu, Yong-Jin Ye,  Jin-Hui Chen, Yu-Gang Ma, Song Zhang, Chen Zhong, Nuclear Science and Techniques {\bf 27}, 86 (2016).
 \urlprefix\url{http://doi/10.1007/s41365-016-0093-7}

\bibitem[{\citenamefont{Lin et~al.}(2001)\citenamefont{Lin, Pal, Ko, Li, and
  Zhang}}]{Lin2001}
\bibinfo{author}{\bibfnamefont{Z.-w.} \bibnamefont{Lin}},
  \bibinfo{author}{\bibfnamefont{S.}~\bibnamefont{Pal}},
  \bibinfo{author}{\bibfnamefont{C.~M.} \bibnamefont{Ko}},
  \bibinfo{author}{\bibfnamefont{B.-A.} \bibnamefont{Li}}, \bibnamefont{and}
  \bibinfo{author}{\bibfnamefont{B.}~\bibnamefont{Zhang}},
  \bibinfo{journal}{Phys. Rev. C} \textbf{\bibinfo{volume}{64}},
  \bibinfo{pages}{011902} (\bibinfo{year}{2001}),
  \urlprefix\url{http://link.aps.org/doi/10.1103/PhysRevC.64.011902}.

\bibitem[{\citenamefont{Voloshin
  et~al.}(2008{\natexlab{b}})\citenamefont{Voloshin, Poskanzer, Tang, and
  Wang}}]{Voloshin2007695}
\bibinfo{author}{\bibfnamefont{S.~A.} \bibnamefont{Voloshin}},
  \bibinfo{author}{\bibfnamefont{A.~M.} \bibnamefont{Poskanzer}},
  \bibinfo{author}{\bibfnamefont{A.}~\bibnamefont{Tang}}, \bibnamefont{and}
  \bibinfo{author}{\bibfnamefont{G.}~\bibnamefont{Wang}},
  \bibinfo{journal}{Phys. Lett.} \textbf{\bibinfo{volume}{B659}},
  \bibinfo{pages}{537} (\bibinfo{year}{2008}{\natexlab{b}}).

\bibitem[{\citenamefont{Qiu and Heinz}(2011)}]{Qiu201184}
\bibinfo{author}{\bibfnamefont{Z.}~\bibnamefont{Qiu}} \bibnamefont{and}
  \bibinfo{author}{\bibfnamefont{U.~W.} \bibnamefont{Heinz}},
  \bibinfo{journal}{Phys. Rev.} \textbf{\bibinfo{volume}{C84}},
  \bibinfo{pages}{024911} (\bibinfo{year}{2011}).

\bibitem[{\citenamefont{Petersen et~al.}(2012)\citenamefont{Petersen, La~Placa,
  and Bass}}]{Petersen2012}
\bibinfo{author}{\bibfnamefont{H.}~\bibnamefont{Petersen}},
  \bibinfo{author}{\bibfnamefont{R.}~\bibnamefont{La~Placa}}, \bibnamefont{and}
  \bibinfo{author}{\bibfnamefont{S.~A.} \bibnamefont{Bass}},
  \bibinfo{journal}{J. Phys.} \textbf{\bibinfo{volume}{G39}},
  \bibinfo{pages}{055102} (\bibinfo{year}{2012}).

\bibitem[{\citenamefont{Voloshin}(2006)}]{Voloshin2006}
\bibinfo{author}{\bibfnamefont{S.~A.} \bibnamefont{Voloshin}}
  (\bibinfo{year}{2006}), \eprint{nucl-th/0606022}.

\bibitem[{\citenamefont{Bhalerao and Ollitrault}(2006)}]{Bhalerao2006}
\bibinfo{author}{\bibfnamefont{R.~S.} \bibnamefont{Bhalerao}} \bibnamefont{and}
  \bibinfo{author}{\bibfnamefont{J.-Y.} \bibnamefont{Ollitrault}},
  \bibinfo{journal}{Phys. Lett.} \textbf{\bibinfo{volume}{B641}},
  \bibinfo{pages}{260} (\bibinfo{year}{2006}), ISSN \bibinfo{issn}{0370-2693},
  \urlprefix\url{http://www.sciencedirect.com/science/article/pii/S0370269306010318}.

\bibitem[{\citenamefont{Bilandzic et~al.}(2011)\citenamefont{Bilandzic,
  Snellings, and Voloshin}}]{Bilandzic2011}
\bibinfo{author}{\bibfnamefont{A.}~\bibnamefont{Bilandzic}},
  \bibinfo{author}{\bibfnamefont{R.}~\bibnamefont{Snellings}},
  \bibnamefont{and} \bibinfo{author}{\bibfnamefont{S.}~\bibnamefont{Voloshin}},
  \bibinfo{journal}{Phys. Rev. C} \textbf{\bibinfo{volume}{83}},
  \bibinfo{pages}{044913} (\bibinfo{year}{2011}),
  \urlprefix\url{http://link.aps.org/doi/10.1103/PhysRevC.83.044913}.

\bibitem[{\citenamefont{Zhou et~al.}(2016)\citenamefont{Zhou, Xiao, Feng, Liu,
  and Snellings}}]{Zhou2015}
\bibinfo{author}{\bibfnamefont{Y.}~\bibnamefont{Zhou}},
  \bibinfo{author}{\bibfnamefont{K.}~\bibnamefont{Xiao}},
  \bibinfo{author}{\bibfnamefont{Z.}~\bibnamefont{Feng}},
  \bibinfo{author}{\bibfnamefont{F.}~\bibnamefont{Liu}}, \bibnamefont{and}
  \bibinfo{author}{\bibfnamefont{R.}~\bibnamefont{Snellings}},
  \bibinfo{journal}{Phys. Rev.} \textbf{\bibinfo{volume}{C93}},
  \bibinfo{pages}{034909} (\bibinfo{year}{2016}).

\bibitem[{\citenamefont{Abelev et~al.}(2014)}]{Abelev2014}
\bibinfo{author}{\bibfnamefont{B.~B.} \bibnamefont{Abelev}}
  \bibnamefont{et~al.} (\bibinfo{collaboration}{ALICE Collaboration}),
  \bibinfo{journal}{Phys. Rev.} \textbf{\bibinfo{volume}{C90}},
  \bibinfo{pages}{054901} (\bibinfo{year}{2014}).

\bibitem[{\citenamefont{Adler et~al.}(2002)\citenamefont{Adler, Ahammed
  et~al.}}]{STAR200266}
\bibinfo{author}{\bibfnamefont{C.}~\bibnamefont{Adler}},
  \bibinfo{author}{\bibfnamefont{Z.}~\bibnamefont{Ahammed}},
  \bibnamefont{et~al.}, \bibinfo{journal}{Phys. Rev. C}
  \textbf{\bibinfo{volume}{66}}, \bibinfo{pages}{034904}
  (\bibinfo{year}{2002}),
  \urlprefix\url{http://link.aps.org/doi/10.1103/PhysRevC.66.034904}.

\bibitem[{\citenamefont{Lacey et~al.}(2011)\citenamefont{Lacey, Wei, Ajitanand,
  and Taranenko}}]{Lacey2010}
\bibinfo{author}{\bibfnamefont{R.~A.} \bibnamefont{Lacey}},
  \bibinfo{author}{\bibfnamefont{R.}~\bibnamefont{Wei}},
  \bibinfo{author}{\bibfnamefont{N.~N.} \bibnamefont{Ajitanand}},
  \bibnamefont{and}
  \bibinfo{author}{\bibfnamefont{A.}~\bibnamefont{Taranenko}},
  \bibinfo{journal}{Phys. Rev.} \textbf{\bibinfo{volume}{C83}},
  \bibinfo{pages}{044902} (\bibinfo{year}{2011}).

\bibitem[{\citenamefont{Bhalerao et~al.}(2011)\citenamefont{Bhalerao, Luzum,
  and Ollitrault}}]{Bhalerao2011}
\bibinfo{author}{\bibfnamefont{R.~S.} \bibnamefont{Bhalerao}},
  \bibinfo{author}{\bibfnamefont{M.}~\bibnamefont{Luzum}}, \bibnamefont{and}
  \bibinfo{author}{\bibfnamefont{J.-Y.} \bibnamefont{Ollitrault}},
  \bibinfo{journal}{Phys. Rev. C} \textbf{\bibinfo{volume}{84}},
  \bibinfo{pages}{054901} (\bibinfo{year}{2011}),
  \urlprefix\url{http://link.aps.org/doi/10.1103/PhysRevC.84.054901}.

\bibitem[{\citenamefont{Yan and Ollitrault}(2014)}]{Yan2013112}
\bibinfo{author}{\bibfnamefont{L.}~\bibnamefont{Yan}} \bibnamefont{and}
  \bibinfo{author}{\bibfnamefont{J.-Y.} \bibnamefont{Ollitrault}},
  \bibinfo{journal}{Phys. Rev. Lett.} \textbf{\bibinfo{volume}{112}},
  \bibinfo{pages}{082301} (\bibinfo{year}{2014}).

\bibitem[{\citenamefont{Yan et~al.}(2014)\citenamefont{Yan, Ollitrault, and
  Poskanzer}}]{Yan201490}
\bibinfo{author}{\bibfnamefont{L.}~\bibnamefont{Yan}},
  \bibinfo{author}{\bibfnamefont{J.-Y.} \bibnamefont{Ollitrault}},
  \bibnamefont{and} \bibinfo{author}{\bibfnamefont{A.~M.}
  \bibnamefont{Poskanzer}}, \bibinfo{journal}{Phys. Rev. C}
  \textbf{\bibinfo{volume}{90}}, \bibinfo{pages}{024903}
  (\bibinfo{year}{2014}),
  \urlprefix\url{http://link.aps.org/doi/10.1103/PhysRevC.90.024903}.

\bibitem[{\citenamefont{Pang et~al.}(2012)\citenamefont{Pang, Wang, and
  Wang}}]{Pang201286}
\bibinfo{author}{\bibfnamefont{L.}~\bibnamefont{Pang}},
  \bibinfo{author}{\bibfnamefont{Q.}~\bibnamefont{Wang}}, \bibnamefont{and}
  \bibinfo{author}{\bibfnamefont{X.-N.} \bibnamefont{Wang}},
  \bibinfo{journal}{Phys. Rev. C} \textbf{\bibinfo{volume}{86}},
  \bibinfo{pages}{024911} (\bibinfo{year}{2012}),
  \urlprefix\url{http://link.aps.org/doi/10.1103/PhysRevC.86.024911}.

\bibitem[{\citenamefont{Pang et~al.}(2016)\citenamefont{Pang, Petersen, Qin,
  Roy, and Wang}}]{Pang201552}
\bibinfo{author}{\bibfnamefont{L.-G.} \bibnamefont{Pang}},
  \bibinfo{author}{\bibfnamefont{H.}~\bibnamefont{Petersen}},
  \bibinfo{author}{\bibfnamefont{G.-Y.} \bibnamefont{Qin}},
  \bibinfo{author}{\bibfnamefont{V.}~\bibnamefont{Roy}}, \bibnamefont{and}
  \bibinfo{author}{\bibfnamefont{X.-N.} \bibnamefont{Wang}},
  \bibinfo{journal}{Eur. Phys. J.} \textbf{\bibinfo{volume}{A52}},
  \bibinfo{pages}{97} (\bibinfo{year}{2016}).

\bibitem[{\citenamefont{Bozek et~al.}(2015)\citenamefont{Bozek, Bzdak, and
  Ma}}]{Bozek2015}
\bibinfo{author}{\bibfnamefont{P.}~\bibnamefont{Bozek}},
  \bibinfo{author}{\bibfnamefont{A.}~\bibnamefont{Bzdak}}, \bibnamefont{and}
  \bibinfo{author}{\bibfnamefont{G.-L.} \bibnamefont{Ma}},
  \bibinfo{journal}{Phys. Lett.} \textbf{\bibinfo{volume}{B748}},
  \bibinfo{pages}{301} (\bibinfo{year}{2015}), \eprint{1503.03655}.

\bibitem[{\citenamefont{Xiao et~al.}(2013)\citenamefont{Xiao, Liu, and
  Wang}}]{Xiao2012}
\bibinfo{author}{\bibfnamefont{K.}~\bibnamefont{Xiao}},
  \bibinfo{author}{\bibfnamefont{F.}~\bibnamefont{Liu}}, \bibnamefont{and}
  \bibinfo{author}{\bibfnamefont{F.}~\bibnamefont{Wang}},
  \bibinfo{journal}{Phys. Rev.} \textbf{\bibinfo{volume}{C87}},
  \bibinfo{pages}{011901} (\bibinfo{year}{2013}).

\bibitem[{\citenamefont{Jia and Huo}(2014)}]{Jia201490}
\bibinfo{author}{\bibfnamefont{J.}~\bibnamefont{Jia}} \bibnamefont{and}
  \bibinfo{author}{\bibfnamefont{P.}~\bibnamefont{Huo}},
  \bibinfo{journal}{Phys. Rev. C} \textbf{\bibinfo{volume}{90}},
  \bibinfo{pages}{034915} (\bibinfo{year}{2014}),
  \urlprefix\url{http://link.aps.org/doi/10.1103/PhysRevC.90.034915}.

\bibitem[{\citenamefont{Back et~al.}(2005)\citenamefont{Back, Baker
  et~al.}}]{PHOBOS200572}
\bibinfo{author}{\bibfnamefont{B.~B.} \bibnamefont{Back}},
  \bibinfo{author}{\bibfnamefont{M.~D.} \bibnamefont{Baker}},
  \bibnamefont{et~al.} (\bibinfo{collaboration}{PHOBOS Collaboration}),
  \bibinfo{journal}{Phys. Rev. C} \textbf{\bibinfo{volume}{72}},
  \bibinfo{pages}{051901} (\bibinfo{year}{2005}),
  \urlprefix\url{http://link.aps.org/doi/10.1103/PhysRevC.72.051901}.

\bibitem[{\citenamefont{Adam et~al.}(2016)}]{Adam2016}
\bibinfo{author}{\bibfnamefont{J.}~\bibnamefont{Adam}} \bibnamefont{et~al.}
  (\bibinfo{collaboration}{ALICE}) (\bibinfo{year}{2016}), \eprint{1605.02035}.

\bibitem[{\citenamefont{Kolb et~al.}(2004)\citenamefont{Kolb, Chen, Greco, and
  Ko}}]{Kolb200469}
\bibinfo{author}{\bibfnamefont{P.~F.} \bibnamefont{Kolb}},
  \bibinfo{author}{\bibfnamefont{L.-W.} \bibnamefont{Chen}},
  \bibinfo{author}{\bibfnamefont{V.}~\bibnamefont{Greco}}, \bibnamefont{and}
  \bibinfo{author}{\bibfnamefont{C.~M.} \bibnamefont{Ko}},
  \bibinfo{journal}{Phys. Rev.} \textbf{\bibinfo{volume}{C69}},
  \bibinfo{pages}{051901} (\bibinfo{year}{2004}).

\bibitem[{\citenamefont{Schulc and Tom‡?ik}(2014)}]{Schulc2014}
\bibinfo{author}{\bibfnamefont{M.}~\bibnamefont{Schulc}} \bibnamefont{and}
  \bibinfo{author}{\bibfnamefont{B.}~\bibnamefont{Tom‡?ik}},
  \bibinfo{journal}{Phys. Rev.} \textbf{\bibinfo{volume}{C90}},
  \bibinfo{pages}{064910} (\bibinfo{year}{2014}).

\bibitem[{\citenamefont{Nie and Ma}(2014{\natexlab{a}})}]{Nie100519}
\bibinfo{author}{\bibfnamefont{M.}~\bibnamefont{Nie}} \bibnamefont{and}
  \bibinfo{author}{\bibfnamefont{G.}~\bibnamefont{Ma}},
  \bibinfo{journal}{Nuclear Techniques (in Chinese)}
  \textbf{\bibinfo{volume}{37}}, \bibinfo{pages}{100519}
  (\bibinfo{year}{2014}{\natexlab{a}}),
  \urlprefix\url{http://www.j.sinap.ac.cn/hejishu/CN/abstract/article_288.shtml}.

\bibitem[{\citenamefont{Khachatryan et~al.}(2015)}]{Khachatryan2015}
\bibinfo{author}{\bibnamefont{Khachatryan}} \bibnamefont{et~al.}
  (\bibinfo{collaboration}{CMS Collaboration}), \bibinfo{journal}{Phys. Rev. C}
  \textbf{\bibinfo{volume}{92}}, \bibinfo{pages}{034911}
  (\bibinfo{year}{2015}),
  \urlprefix\url{http://link.aps.org/doi/10.1103/PhysRevC.92.034911}.

\bibitem[{\citenamefont{Lacey et~al.}(2014)\citenamefont{Lacey, Taranenko, Jia,
  Reynolds, Ajitanand, Alexander, Gu, and Mwai}}]{Lacey2014112}
\bibinfo{author}{\bibfnamefont{R.~A.} \bibnamefont{Lacey}},
  \bibinfo{author}{\bibfnamefont{A.}~\bibnamefont{Taranenko}},
  \bibinfo{author}{\bibfnamefont{J.}~\bibnamefont{Jia}},
  \bibinfo{author}{\bibfnamefont{D.}~\bibnamefont{Reynolds}},
  \bibinfo{author}{\bibfnamefont{N.~N.} \bibnamefont{Ajitanand}},
  \bibinfo{author}{\bibfnamefont{J.~M.} \bibnamefont{Alexander}},
  \bibinfo{author}{\bibfnamefont{Y.}~\bibnamefont{Gu}}, \bibnamefont{and}
  \bibinfo{author}{\bibfnamefont{A.}~\bibnamefont{Mwai}},
  \bibinfo{journal}{Phys. Rev. Lett.} \textbf{\bibinfo{volume}{112}},
  \bibinfo{pages}{082302} (\bibinfo{year}{2014}),
  \urlprefix\url{http://link.aps.org/doi/10.1103/PhysRevLett.112.082302}.

\bibitem[{\citenamefont{Gardim et~al.}(2012)\citenamefont{Gardim, Grassi,
  Luzum, and Ollitrault}}]{Gardim2012}
\bibinfo{author}{\bibfnamefont{F.~G.} \bibnamefont{Gardim}},
  \bibinfo{author}{\bibfnamefont{F.}~\bibnamefont{Grassi}},
  \bibinfo{author}{\bibfnamefont{M.}~\bibnamefont{Luzum}}, \bibnamefont{and}
  \bibinfo{author}{\bibfnamefont{J.-Y.} \bibnamefont{Ollitrault}},
  \bibinfo{journal}{Phys. Rev. C} \textbf{\bibinfo{volume}{85}},
  \bibinfo{pages}{024908} (\bibinfo{year}{2012}),
  \urlprefix\url{http://link.aps.org/doi/10.1103/PhysRevC.85.024908}.

\bibitem[{\citenamefont{Jia}(2014)}]{Jia201441}
\bibinfo{author}{\bibfnamefont{J.}~\bibnamefont{Jia}}, \bibinfo{journal}{J.
  Phys.} \textbf{\bibinfo{volume}{G41}}, \bibinfo{pages}{124003}
  (\bibinfo{year}{2014}).

\bibitem[{\citenamefont{Zhang and Liao}(2014)}]{Zhang201489}
\bibinfo{author}{\bibfnamefont{X.}~\bibnamefont{Zhang}} \bibnamefont{and}
  \bibinfo{author}{\bibfnamefont{J.}~\bibnamefont{Liao}},
  \bibinfo{journal}{Phys. Rev. C} \textbf{\bibinfo{volume}{89}},
  \bibinfo{pages}{014907} (\bibinfo{year}{2014}),
  \urlprefix\url{http://link.aps.org/doi/10.1103/PhysRevC.89.014907}.

\bibitem[{\citenamefont{Zhang and Liao}(2012)}]{Zhang2012713}
\bibinfo{author}{\bibfnamefont{X.-L.} \bibnamefont{Zhang}} \bibnamefont{and}
  \bibinfo{author}{\bibfnamefont{J.-F.} \bibnamefont{Liao}},
  \bibinfo{journal}{Phys. Lett.} \textbf{\bibinfo{volume}{B713}},
  \bibinfo{pages}{35} (\bibinfo{year}{2012}).

\bibitem[{\citenamefont{Nie and Ma}(2014{\natexlab{b}})}]{Nie2014}
\bibinfo{author}{\bibfnamefont{M.}~\bibnamefont{Nie}} \bibnamefont{and}
  \bibinfo{author}{\bibfnamefont{G. L.}~\bibnamefont{Ma}}, \bibinfo{journal}{Phys.
  Rev.} \textbf{\bibinfo{volume}{C90}}, \bibinfo{pages}{014907}
  (\bibinfo{year}{2014}{\natexlab{b}}).
  \urlprefix\url{http://dx.doi.org/10.1103/PhysRevC.90.014907}.

\end{thebibliography}


\end{document}